\documentclass[%
 aip,
 amsmath,amssymb,
 reprint,%
]{revtex4-1}

\usepackage{graphicx}
\usepackage{dcolumn}
\usepackage{bm}

\usepackage[utf8]{inputenc}
\usepackage[T1]{fontenc}
\usepackage{mathptmx}
\usepackage{etoolbox}

\usepackage{xcolor,siunitx}
\usepackage{enumitem}
\usepackage{threeparttable}
\usepackage[caption=false]{subfig}
\usepackage{multirow}
\usepackage{tabularx}

\usepackage{hyperref}
\hypersetup{bookmarks,
    colorlinks=true,
    citecolor=red,
    linkcolor=blue,
    urlcolor=cyan,
    pagebackref=true}

\makeatletter
\def\@email#1#2{%
 \endgroup
 \patchcmd{\titleblock@produce}
  {\frontmatter@RRAPformat}
  {\frontmatter@RRAPformat{\produce@RRAP{*#1\href{mailto:#2}{#2}}}\frontmatter@RRAPformat}
  {}{}
}%
\makeatother
\begin{document}

\preprint{AIP/123-QED}

\title[]{Reconciling spectroscopy with dynamics in global potential energy surfaces: the case of the astrophysically relevant SiC$_{\bf 2}$}
\author{C.~M.~R.~Rocha$^{*}$}
\email{romerorocha@strw.leidenuniv.nl.}
\author{H.~Linnartz}%
\affiliation{ 
Laboratory for Astrophysics, Leiden Observatory, Leiden University, P.O. Box 9513, NL-2300 RA Leiden, The Netherlands}%
\author{A.~J.~C.~Varandas}
\affiliation{%
School of Physics and Physical Engineering, Qufu Normal University, 273165 Qufu, China
}%
\affiliation{%
Department of Physics, Universidade Federal do Esp\'{i}rito Santo,
29075-910 Vit\'{o}ria, Brazil
}%
\affiliation{%
Department of Chemistry, and Chemistry Centre, University of Coimbra,
3004-535 Coimbra, Portugal
}%

\date{\today}

\begin{abstract}
SiC$_2$ is a fascinating molecule 
due to its unusual bonding and astrophysical importance. 
In this work, we 
report the first global potential energy surface (PES) for ground-state SiC$_2$ using the combined-hyperbolic-inverse-power-representation (CHIPR) method and 
accurate \emph{ab initio} energies. {The calibration grid data is obtained via a general dual-level protocol developed afresh herein that entails both coupled-cluster and multireference configuration interaction energies jointly extrapolated to the complete basis set limit. Such an approach is specially devised to recover much of the spectroscopy from the PES, while still permitting a proper fragmentation of the system to allow for reaction dynamics studies}. 
Besides describing accurately the valence strongly-bound region that 
includes both the cyclic global minimum and isomerization barriers, 
the final analytic PES form is shown to properly reproduce dissociation energies, diatomic potentials, and long-range interactions at all asymptotic channels, in addition to naturally reflect the correct permutational symmetry of the potential. 
Bound vibrational state calculations have been carried out, 
unveiling an excellent match of the available experimental data 
on $c$-$\mathrm{SiC}_{2}(^{1}A_1)$. To further exploit the global nature of 
the PES, exploratory quasi-classical trajectory calculations 
for the endothermic $\mathrm{C_{2}\!+\!Si}\rightarrow\mathrm{SiC\!+\!C}$ reaction are also performed, yielding thermalized rate coefficients for 
temperatures up to $5000\,\si{\kelvin}$. The results hint for the prominence of this reaction in the innermost layers of the circumstellar
envelopes around carbon-rich stars, thence conceivably playing therein
a key contribution to the gas-phase formation of SiC, and eventually, solid SiC dust.
\end{abstract}

\maketitle

\section{Introduction}\label{sec:intro}
Silicon dicarbide, SiC$_2$, has enjoyed a great deal of 
attention 
for its applications in   astrochemistry~\cite{THA84:L45,NYM93:377,GEN97:819,SAR000:103,MOR004:1196,CER010:L8,KOK011:17,MUL012:50,GOB017:117,MAS018:A29,CER018:A4,MCC019:7,AGU020:A59,SHA020:1869}:
\begin{enumerate}[label=(\roman*).]
 \item{Its most stable cyclic $C_{2v}$ isomer, $c$-$\mathrm{SiC}_{2}(^{1}A_1)$, was the first molecular ring identified in the interstellar medium~\cite{THA84:L45}.}
 \item{Its Merrill-Sanford band system ($\tilde{A}^{1}B_2$--$\tilde{X}^{1}A_1$ electronic 
 transition) near 5000\,{\AA} was first observed in the optical absorption spectra of evolved stars, and continues to be a particularly valuable astronomical probe of stellar atmospheres~\cite{SAR000:103,MOR004:1196}.}
 \item{Besides rovibronic transitions, the pure rotational signatures of both main ($^{28}$Si$^{12}$C$_2$) and singly-substituted isotopologues ($^{29}$SiC$_2$, $^{30}$SiC$_2$, and $^{28}$Si$^{13}$CC) of $c$-$\mathrm{SiC}_{2}$ have 
 been identified in several astrophysical sources~\cite{THA84:L45,NYM93:377,CER010:L8,KOK011:17,MUL012:50,MAS018:A29,CER018:A4} and serve as sensitive molecular diagnostic tools for probing 
 the chemical and physical
 conditions of the regions in which they reside~\cite{THA84:L45,CER010:L8}.}
 \item{Together with SiC and Si$_{2}$C parent molecules, 
 $c$-$\mathrm{SiC}_{2}$ is ranked among the most likely  
 gas-phase precursors leading to the formation of 
 SiC dust grains 
 in the inner envelopes of late-type carbon-rich stars~\cite{GOB017:117,MAS018:A29,MCC019:7,AGU020:A59}.}
\end{enumerate}

Apart from its intrinsic interest in an astronomical context, 
SiC$_2$ is also a fascinating molecule from a chemical viewpoint 
owing to its unique structure and 
dynamics~\cite{ROS94:4110,NIE97:1195}. 
Previous laboratory~\cite{MIC84:3556,BUT91:1,ROS94:4110} and 
quantum mechanical studies~\cite{NIE97:1195,KEN003:7353,FOR015:13,KOP016:2395} 
jointly provided ample evidence that its lowest energy $C_{2v}$ minimum (as definitively assigned by Michalopoulos~\emph{et al.}~\cite{MIC84:3556}) 
has an exceedingly flat potential energy surface (PES) along the internal rotation of the C$_2$ moiety within the molecule~\cite{MIC84:3556,BUT91:1,ROS94:4110,NIE97:1195,KEN003:7353,FOR015:13,KOP016:2395}. Such an untypical, nondirectional Si--C$_{2}$ bonding in $c$-$\mathrm{SiC}_{2}$ (with reportedly high ionic character) has been classified~\cite{NIE97:1195,ODD85:1702} as polytopic~\cite{CLE73:2460} 
in nature, and hence characterized by the nearly-free circumnavigation 
of Si about C$_{2}$~\cite{CLE73:2460}. Indeed, the expectedly low energy difference between $c$-$\mathrm{SiC}_{2}$ and the linear $C_{\infty v}$ ($\ell$-$\mathrm{SiCC}$) saddle-point structure was first confirmed experimentally by Ross~\emph{et al.}~\cite{ROS94:4110} as being only $\sim\!1883\,\mathrm{cm^{-1}}$. 
{ Clearly, like in C$_3$~\cite{ROC015:074302,ROC018:36}, the expected high vibrational state populations and their delocalization over large regions of the PES make $c$-SiC$_2$'s intramolecular motion lying at the borderlines of spectroscopy and chemical dynamics.}

The conclusions drawn 
from these early experimental works by Michalopoulos~\emph{et al.}~\cite{MIC84:3556}~and~Ross~\emph{et al.}~\cite{ROS94:4110} motivated a plethora of detailed spectroscopic studies on $c$-$\mathrm{SiC}_{2}$ aiming to further characterize its spectral signatures in both  microwave~\cite{GOT89:L29,KOK011:17,MUL012:50,CER018:A4,MCC019:7}, infrared~\cite{SHE85:4788,PRE90:5424,IZU94:1371} and optical~\cite{BUT91:1,ZHA020:111306} regions; for a comprehensive review, see Ref.~\citenum{ZHA020:111306} and references therein. 

From the theoretical perspective, several concurring investigations  
were also ignited towards unraveling the SiC$_2$'s unusual polytopic bonding nature and its large-amplitude dynamics~\cite{GRE83:895,ODD85:1702,ARU95:71,NIE97:1195,ZHA98:31,KEN003:7353,FOR015:13,KOP016:2395}; for a complete account of the earlier theoretical literature the reader is addressed to Refs.~\citenum{NIE97:1195}~and~\citenum{KOP016:2395}. 
In the most recent studies by Fortenberry~\emph{et al.}~\cite{FOR015:13}~and~Koput~\cite{KOP016:2395}, 
special emphasis were put into the characterization of the $c$-$\mathrm{SiC}_{2}$'s local PES using state-of-the-art \emph{ab initio} composite methods. 
By relying on the so-called CcCR protocol~\cite{FOR015:13}, Fortenberry~\emph{et al.} reported a near-equilibrium quartic force field (QFF) for silicon dicarbide; the QFF was based on CCSD(T) energies extrapolated to the complete basis set (CBS) limit, 
augmented by additive corrections due to core-electron correlation and relativistic effects~\cite{FOR015:13}. Using standard vibrational perturbational theory (VPT2), the CcCR QFF has shown to reproduce the $c$-$\mathrm{SiC}_{2}$'s stretching fundamentals ($\nu_1$ and $\nu_2$) 
to within $5\,\mathrm{cm^{-1}}$ of experiment~\cite{BUT91:1}, 
whereas larger deviations of up to $21\,\mathrm{cm^{-1}}$ have been found for the $\nu_3$ (C$_2$ hindered rotation) mode~\cite{FOR015:13}. As noted by Nielsen~\emph{et al.}~\cite{NIE97:1195}~and~Koput~\cite{KOP016:2395}, 
this is not surprising given the inherent deficiencies of VPT2 in properly describing such highly-anharmonic, large-amplitude pinwheel dynamics of $c$-$\mathrm{SiC}_{2}$. In the most sophisticated theoretical study to date by Jacek Koput~\cite{KOP016:2395}, a more extended PES (hereinafter referred to as JK PES) was reported that describes locally 
not only $c$-$\mathrm{SiC}_{2}$ but also 
the $\ell$-$\mathrm{SiCC}$ saddle-point, in addition to the minimum energy path connecting them; the calibration  
data set included \emph{ab initio} CCSD(T)-F12b/cc-pCVQZ-F12 energies 
additively corrected for higher-order valence-electron correlation beyond CCSD(T) and scalar relativistic effects~\cite{KOP016:2395}. Its barrier to linearity  
was predicted to be $1782\,\mathrm{cm^{-1}}$ which is lower than the previous high-level \emph{ab initio} estimates by Nielsen~\emph{et al.}~\cite{NIE97:1195} ($2030\,\mathrm{cm^{-1}}$), and Kenny~\emph{et al.}~\cite{KEN003:7353} ($2210\,\mathrm{cm^{-1}}$), but closer to the experimental value~\cite{ROS94:4110} of $1883\,\mathrm{cm^{-1}}$. Based on a variational approach, Koput~\cite{KOP016:2395} also performed bound-state calculations on his final potential; the results have shown that the JK PES is capable 
of reproducing the $c$-$\mathrm{SiC}_{2}$'s 
experimental vibrational term values   
reported by Ross~\emph{et al.}~\cite{ROS94:4110} 
with a root-mean-square deviation (rmsd) of $\sim\!5\,\mathrm{cm^{-1}}$. 

Clearly, all the above distinctive features of $\mathrm{SiC}_{2}$ make it 
a challenging testing ground 
for any theoretical methodological development. 
Moreover, the expected implications its unique spectroscopy and
reaction dynamics might have in molecular astrophysics, render this 
molecule a tempting target for further studies. 
{As noted above, previous theoretical studies were mainly  
concerned with the determination of locally valid spectroscopic potentials for $c$-$\mathrm{SiC}_{2}$~\cite{NIE97:1195,FOR015:13,KOP016:2395} and 
there is not as yet a global PES for the title system that is capable of  
accurately describing both its valence and dissociation features at once.}  
In this work, we delve deeper into the silicon dicarbide saga~\cite{NIE97:1195} and 
provide {for the first time} such a form for ground-state $\mathrm{SiC}_{2}$. 
{To allow for both bound-state and reaction dynamics calculations,} the PES will be based on an 
accurate \emph{ab initio} 
protocol 
that incorporates 
the best of two worlds:~coupled-cluster [CCSD(T)] and multireference configuration interaction [MRCI(Q)] energies jointly extrapolated to the CBS limit. 
For the analytical modeling, we employ 
the Combined-Hyperbolic-Inverse-Power-Representation (CHIPR) 
method~\cite{VAR013:054120,VAR013:408,ROC020:106913,ROC021:107556} as implemented in 
the CHIPR-4.0 program~\cite{ROC021:107556}. 
The quality of the final potential is further judged via both 
spectroscopic and exploratory reaction dynamics calculations. 


\section{Methodology}\label{sec:method}
\subsection{\emph{Ab initio} calculations}\label{subsec:abinitioextrap}
\begin{figure}
\centering
\includegraphics[angle=0,width=1\linewidth]{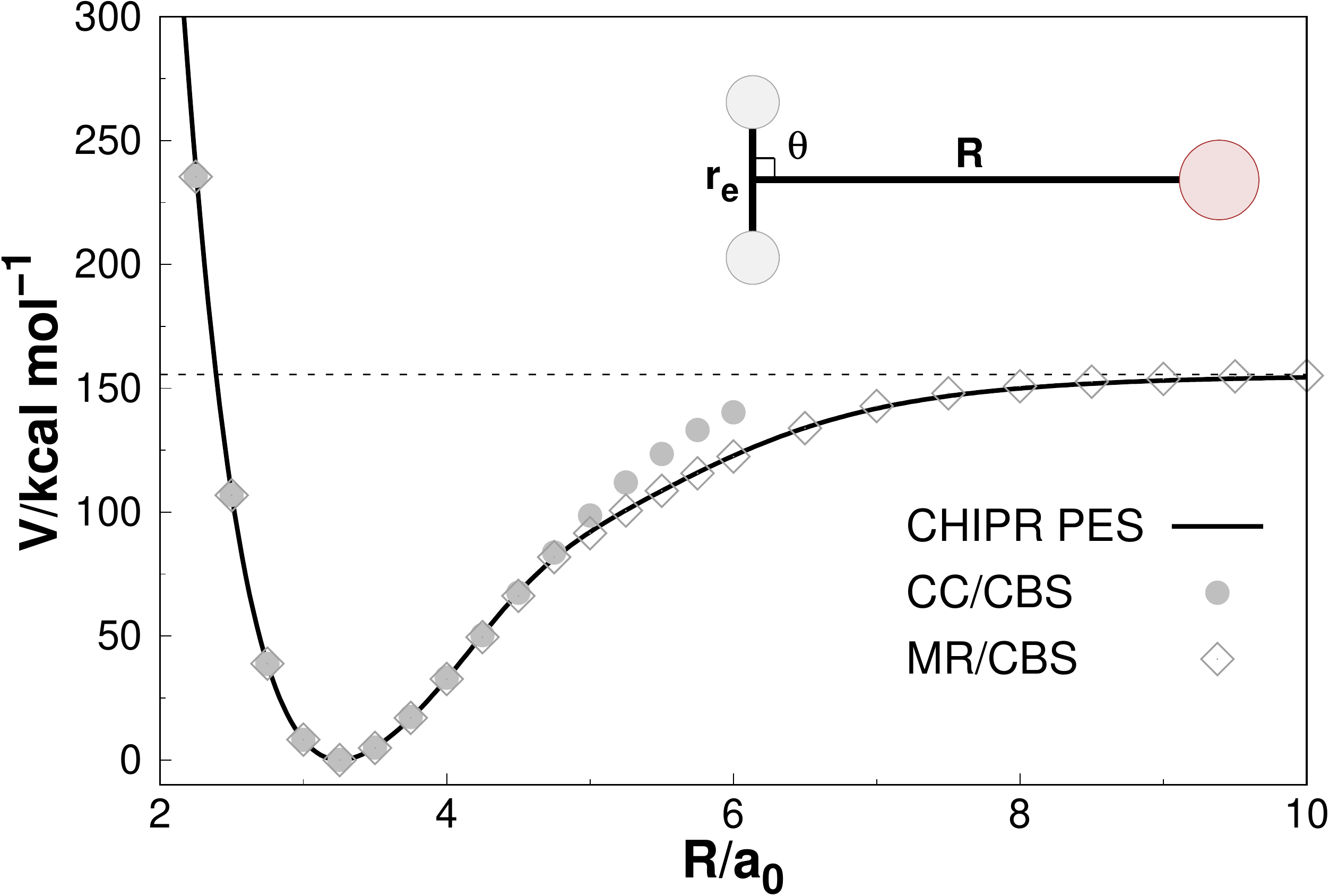}
\caption{\footnotesize Extrapolated CCSD(T)~[CC/CBS] and MRCI(Q)~[MR/CBS] energies 
for a cut along the perpendicular {{($\theta\!=\!90\si{\degree}$)}} approach of a Si atom into the C$_{2}$ diatomic
with $r_{e}\!=\!2.40133\,a_{0}$. The corresponding final CHIPR PES is also shown for comparison. The zero of energy corresponds 
to the T-shaped ($C_{2v}$) global minimum at $R\!=\!3.25350\,a_{0}$.}
\label{fig:cut}
\end{figure}

All electronic structure calculations have been done with~{MOLPRO}~\citep{MOLPRO}. To ensure an accurate description of both valence and long-range features of the PES, 
the full set of \emph{ab initio} grid points were herein generated 
using a combination~\cite{GAL009:14424} of CCSD(T)~\cite{KNO93:5219,PIE007:63,BAR007:291}~(CC for brevity) and MRCI(Q)~\cite{SZA012:108}~(MR) levels of theory. The first is specially devised to improve the spectroscopy of the global minimum and is limited 
[due to the well-known~\cite{GAL009:14424,SZA012:108} erratic behaviour of such single-reference method for stretched bond distances (Figure~\ref{fig:cut})] 
to a small region of the PES near 
the $c$-$\mathrm{SiC}_{2}$/$\ell$-$\mathrm{SiCC}$ stationary points.  
The MR set is in turn responsible to cover the bulk of the PES~\cite{SZA012:108}, being restricted to sample the fragmentation region {and geometries with high $T_{1}$ and $D_{1}$ diagnostics~\cite{LEE89:199,JAN98:423} [\emph{e.g.}, those characterized by 
larger $\mathrm{C}\!-\!\mathrm{C}$ bond distances, away from the equilibrium region; see Figure~\ref{fig:hyper}(a) later].} {Both data sets were subsequently  
extrapolated} to 
the CBS limit~\cite{VAR018:177,VAR020:2030001} (see below). The AV$X$Z\,($X\!=\!T,Q,5$) basis sets of Dunning and co-workers~\cite{DUN89:1007,KEN92:6796} including  
additional tight-$d$ functions ($+d$) for the Si atom~\cite{MOLPRO} 
were employed throughout.  

At each selected geometry $\mathbf{R}$, 
the CC/CBS energy was defined as~\cite{ROC019:24406}
\begin{equation}\label{eq:cccbs}
E_{\infty}^{\mathrm{CC}}(\mathbf{R})=E_{\infty}^{\mathrm{HF}}(\mathbf{R})
+E_{\infty}^{\mathrm{cor}}(\mathbf{R}), 
\end{equation}
where $E_{\infty}^{\mathrm{HF}}$ and $E_{\infty}^{\mathrm{cor}}$ are the extrapolated 
HF and CC correlation (cor) components. 
In Eq.~(\ref{eq:cccbs}), $E_{\infty}^{\mathrm{HF}}$ is obtained via a two-point extrapolation  
protocol~\cite{PAN016:261}
\begin{equation}\label{eq:hfcbs}
E_{X}^{\mathrm{HF}}(\mathbf{R})=E_{\infty}^{\mathrm{HF}}(\mathbf{R})+Ae^{-\beta x}
\end{equation}
where $x\!=\!q\,(3.87),\,p\,(5.07)$ are hierarchical 
numbers~\cite{VAR014:224113,PAN015:90} that parallel the traditional $X\!=\!Q,5$ cardinal ones, $\beta\!=\!1.62$, and $E_{\infty}^{\mathrm{HF}}$ and $A$ are parameters   
to be calibrated from the raw RHF/AV$X$Z ($X\!=\!Q,5$) 
energies~\cite{PAN016:261}. 
In turn, $E_{\infty}^{\mathrm{cor}}$ is obtained using the inverse-power formula~\cite{VAR014:224113}
\begin{equation}\label{eq:corcbs}
E_{X}^{\mathrm{cor}}(\mathbf{R})=E_{\infty}^{\mathrm{cor}}(\mathbf{R})+\frac{A_{3}}{x^3},
\end{equation}
where $q\,(3.68),\,p\,(4.71)$ are CC-type $x$ numbers~\cite{VAR014:224113}, with 
$E_{\infty}^{\mathrm{cor}}$ and $A_{3}$ calibrated from the raw CC/AV$X$Z ($X\!=\!Q,5$) cor energies. 

Similarly to Eq.~(\ref{eq:cccbs}), the CBS extrapolations of MR energies 
were performed individually for the non-dynamical (CAS) and dynamical (dc) correlations~\cite{ROC019:8154}
\begin{equation}\label{eq:cascbs}
E_{\infty}^{\mathrm{MR}}(\mathbf{R})=E_{\infty}^{\mathrm{CAS}}(\mathbf{R})
+E_{\infty}^{\mathrm{dc}}(\mathbf{R}), 
\end{equation}
where $E_{\infty}^{\mathrm{CAS}}$ is obtained using Eq.~(\ref{eq:hfcbs}) 
but with CASSCF(12,12)/AV$X$Z ($X\!=\!T,Q$) raw energies~\cite{PAN016:261}   
and $E_{\infty}^{\mathrm{dc}}$ is given by the two-point law~\cite{VAR007:244105}
\begin{equation}\label{eq:dccbs}
E_{X}^{\mathrm{dc}}(\mathbf{R})=E_{\infty}^{\mathrm{dc}}(\mathbf{R})+\frac{A_{3}}{(X\!-\!3/8)^{3}}+
\frac{A_{5}^{\rm o}+c{A_{3}}^{5/4}}{(X\!-\!3/8)^{5}}; 
\end{equation}
here, $A_{5}^{\rm o}$ and $c$ are universal-type parameters~\cite{VAR007:244105}, and $E_{\infty}^{\mathrm{dc}}$ and $A_{3}$ are   
obtained from the raw MRCI(Q)/AV$X$Z ($X\!=\!T,Q$) dc energies. 
{The full-valence CASSCF active space includes the 3s- and 3p-like orbitals of Si 
and the 2s- and 2p-like orbitals of the C atoms. 
Note that, in the CC calculations, core correlation was not taken into account as this would imply, for reasons of 
consistency between both data sets, the consideration of such effects 
also at MR level, making the task of obtaining the global PES 
computationally unaffordable with current available resources. Thus, in all CC and MR calculations, 
only the valence electrons were correlated, with the 2s- and 2p-like orbitals of Si being included into the core.}

Using the above dual-level CC/MR CBS protocol, a total of 3682 symmetry unique  
points (1144 and 2538 at CC/CBS and MR/CBS levels, respectively) 
have been selected to map all relevant regions of the ground-state PES of SiC$_{2}$ using atom--diatom Jacobi coordinates~\cite{MUR84:MPEF} ($r$, $R$, and $\theta$ {{in Figure~\ref{fig:cut}}});   
the ranges are 
$2.0\!\leq\!R/a_{0}\!\leq\!15.0$, $2.0\!\leq\!r/a_{0}\!\leq\!3.5$, and $0.0\!\leq\!\theta/\mathrm{deg}\!\leq\!90.0$ for the Si--C$_{2}$ channel and 
$1.2\!\leq\!R/a_{0}\!\leq\!15.0$, $2.8\!\leq\!r/a_{0}\!\leq\!4.3$, and $0.0\!\leq\!\theta/\mathrm{deg}\!\leq\!180.0$ for C--SiC interactions. 
{Recall that, in partitioning the nuclear configuration space, the CC/CBS data set was chosen to 
cover only a limited region around the global minimum (including $\ell$-$\mathrm{SiCC}$), 
while the MR/CC method was utilized elsewhere.}
Note that the corresponding C$_{2}$ and SiC curves were obtained solely at the MR/CBS level by making atom-diatom calculations with the Si and 
C atoms $50\,a_{0}$ far apart, varying the diatomic internuclear distance only; 
the total number of computed points for each curve amounts to $\sim\!63$ and covers the coordinate range of $1.0\!\leq\!r/a_{0}\!\leq\!50$. 
The reader is addressed to Figure~\ref{fig:hyper}(a)~and~Figures~S1~and~S2 of the Supplementary Material (SM) to assess the full set of \emph{ab initio} grid points.  

{Finally, it should be noted that, while the use of larger basis sets would be desirable in estimating the CBS limits in Eqs.~(\ref{eq:cccbs})-(\ref{eq:dccbs}), 
preliminary test calculations have shown that the associated computational cost to obtain the full global PES would be nearly three times as high 
if the cardinal numbers in the above extrapolation formulas were increased by one unit. 
Because our proposed MR/CBS($T,Q$) and CC/CBS($Q,5$) protocols have already shown 
excellent performances when assessed against benchmark CBS 
energies~\cite{VAR020:2030001,VAR018:177,PAN016:261,VAR014:224113,PAN015:90,VAR007:244105},
we deeemed that there was no reason to extend the one-particle bases further. 
}

\subsection{Calibration of CHIPR PES}\label{subsec:chipr}
\begin{table}
\centering
\caption{\footnotesize Stratified root-mean-square 
deviations (in $\rm kcal\,mol^{-1}$) of the final PES.}
\label{tab:rmsd}
\begin{ruledtabular}
\begin{threeparttable}
\begin{tabular}{ddccc
}
{\rm Energy\tnote{a}} & {N\tnote{b}} & {max.~dev.\tnote{c}} & {rmsd} & {$N_{>\,\rm rmsd}$\tnote{d}} \\[0.5ex] 
\hline \\[-2.25ex]
  50 &  804 & 1.5 & 0.2 & 156 \\
 100 & 1102 & 3.0 & 0.3 & 179 \\
 150 & 2088 & 3.5 & 0.5 & 249 \\
 200 & 2723 & 5.0 & 0.8 & 453 \\
 250 & 3430 & 5.1 & 0.9 & 588 \\
 500 & 3578 & 5.1 & 0.9 & 658 \\
1200 & 3682 & 5.1 & 0.9 & 680 \\
\end{tabular}
\begin{tablenotes}[flushleft]
  \item[a]{{\footnotesize {The units of energy are $\rm kcal\,mol^{-1}$. 
   Energy strata defined with respect to the $C_{2v}$ absolute minimum of SiC$_{2}$: $-364.993433\,\rm E_{h}$ at CCSD(T)/CBS level. 
   Its relative energy (as predicted from the PES with respect to the infinitely separated C+C+Si atoms) is $-0.474269\,\rm E_{h}$}.}}
  \item[b]{{\footnotesize Number of calculated points up to indicated energy range}.}
  \item[c]{{\footnotesize Maximum deviation up to indicated energy range}.}
  \item[d]{{\footnotesize Number of calculated points with an energy deviation larger than 
  the rmsd}.}
\end{tablenotes}
\end{threeparttable}
\end{ruledtabular}
\end{table}
\begin{figure}
\centering
\includegraphics[angle=0,width=1\linewidth]{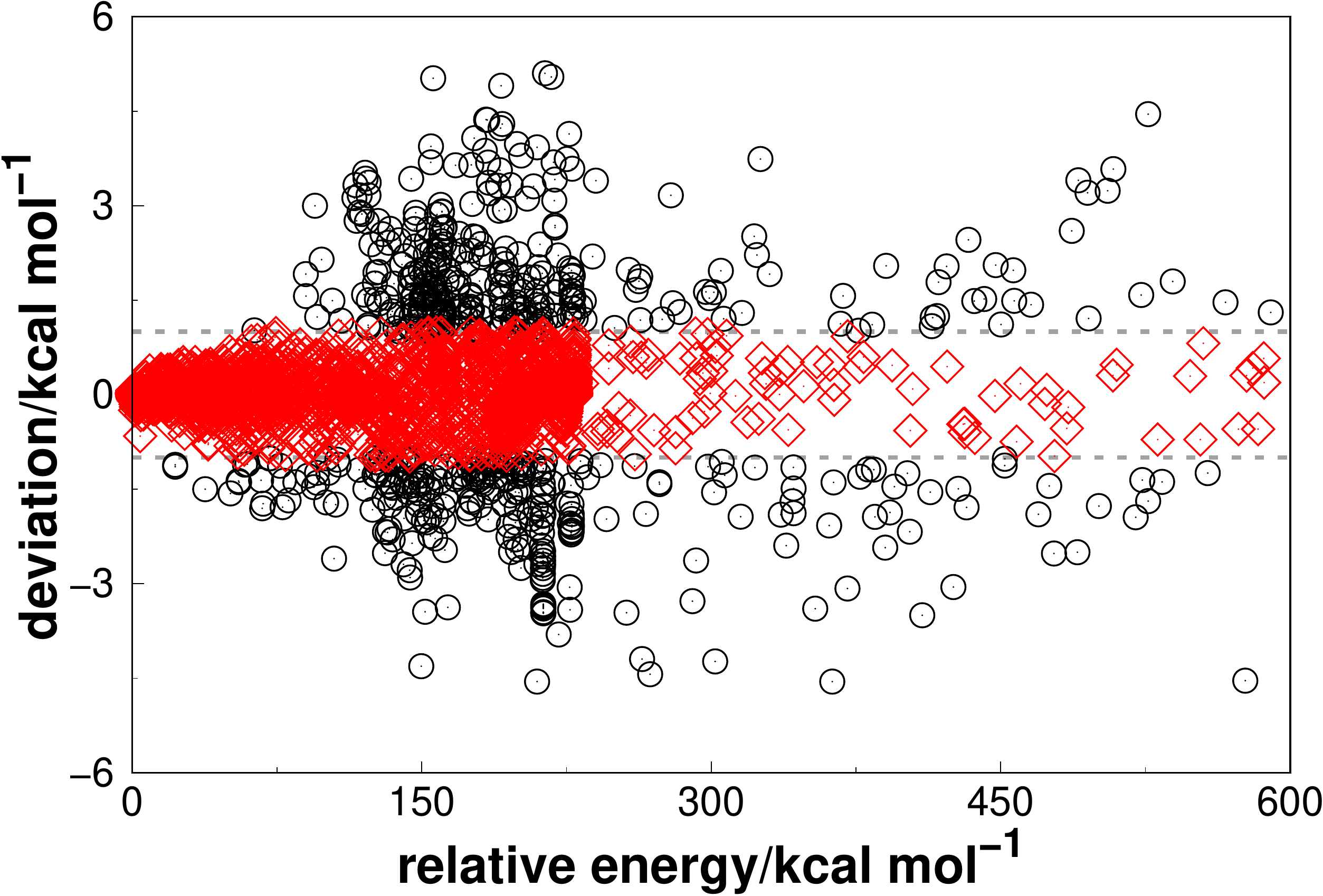}
\caption{\footnotesize {{Scatter plot of deviations between fitted 
[using the CHIPR model function of Eq.~(\ref{eq:pes})] and calculated 
\emph{ab initio} energies as a function of the total energy. 
In the x-axis, the zero is set relative to the $C_{2v}$ global minimum of SiC$_{2}$}}. 
Points fitted with chemical accuracy {{($|\text{deviation}|\!\leq\!1\rm kcal\,mol^{-1}$)}} are represented in red.}
\label{fig:devdist}
\end{figure}

Within the CHIPR~\cite{VAR013:054120,VAR013:408,ROC020:106913,ROC021:107556} formalism, 
the global adiabatic PES of ground-state $\mathrm{SiC}_{2}(^{1}A')$ assumes the following many-body expansion form~\cite{MUR84:MPEF}
\begin{align}\label{eq:pes}
V(\mathbf{R})=&V^{(2)}_{\mathrm{C_{2}}}(R_1)+V^{(2)}_{\mathrm{SiC}}(R_2)+V^{(2)}_{\mathrm{SiC}}(R_3) \nonumber \\
&+V^{(3)}_{\mathrm{SiC_{2}}}(\mathbf{R}),
\end{align}
where the $V^{(2)}$'s represent the diatomic (two-body) potentials  
of $\mathrm{C_{2}}(a\,^{3}\Pi_{u})$ and $\mathrm{SiC}(X\,^{3}\Pi)$ and 
$V^{(3)}$ is the 
three-body term; $\mathbf{R}\!=\!\{R_1,R_2,R_3\}$ is the set of interatomic separations, 
with the energy zero set to 
the infinitely separated $\mathrm{C}(^{3}P)\!+\!\mathrm{C}(^{3}P)\!+\!\mathrm{Si}(^{3}P)$ atoms. 
{ As Eq.~(\ref{eq:pes}) indicates, our analytic CHIPR PES dissociates adiabatically into  
$\mathrm{C_{2}}(a\,^{3}\Pi_{u})\!+\!\mathrm{Si}(^{3}P)$ and $\mathrm{SiC}(X\,^{3}\Pi)\!+\!\mathrm{C}(^{3}P)$, 
thence modeling only the lowest electronic singlet state of $\mathrm{SiC}_{2}$ correlating 
to such open shell fragments; this is warranted by including in Eq.~(\ref{eq:pes}) the proper diatomic 
two-body terms and ensuring that $V^{(3)}$ naturally vanishes for large interatomic separations~\cite{ROC020:106913,ROC021:107556}.
Note that, similarly to 
$\mathrm{C}_{3}$~\cite{ROC015:074302,ROC018:36}, 
the ground-state singlet PES of $\mathrm{SiC}_{2}$ does not dissociate adiabatically into ground-state 
$\mathrm{C_{2}}(X\,^{1}\Sigma_{g}^{+})\!+\!\mathrm{Si}(^{3}P)$ fragments which, according to spin-correlation rules~\cite{HER66:MSMSESESPM}, correlate with the triplet manifold of $\mathrm{SiC}_{2}$ states; see~Figure~S3 for further details. Note further that the spin-allowed $\mathrm{C_{2}}(X\,^{1}\Sigma_{g}^{+})\!+\!\mathrm{Si}(^{1}D)$ channel lies~\cite{NIST_ASD,HER79:MSMSCDM} $\approx\!16\,\rm kcal\,mol^{-1}$ 
above the $\mathrm{C_{2}}(a\,^{3}\Pi_{u})\!+\!\mathrm{Si}(^{3}P)$ asymptote and correlates with excited singlet PESs~\cite{ROC015:074302,ROC018:36}.}  
\begin{figure*}
\captionsetup[subfigure]{labelformat=empty}
\centering
\subfloat{{\includegraphics[width=0.495\linewidth]{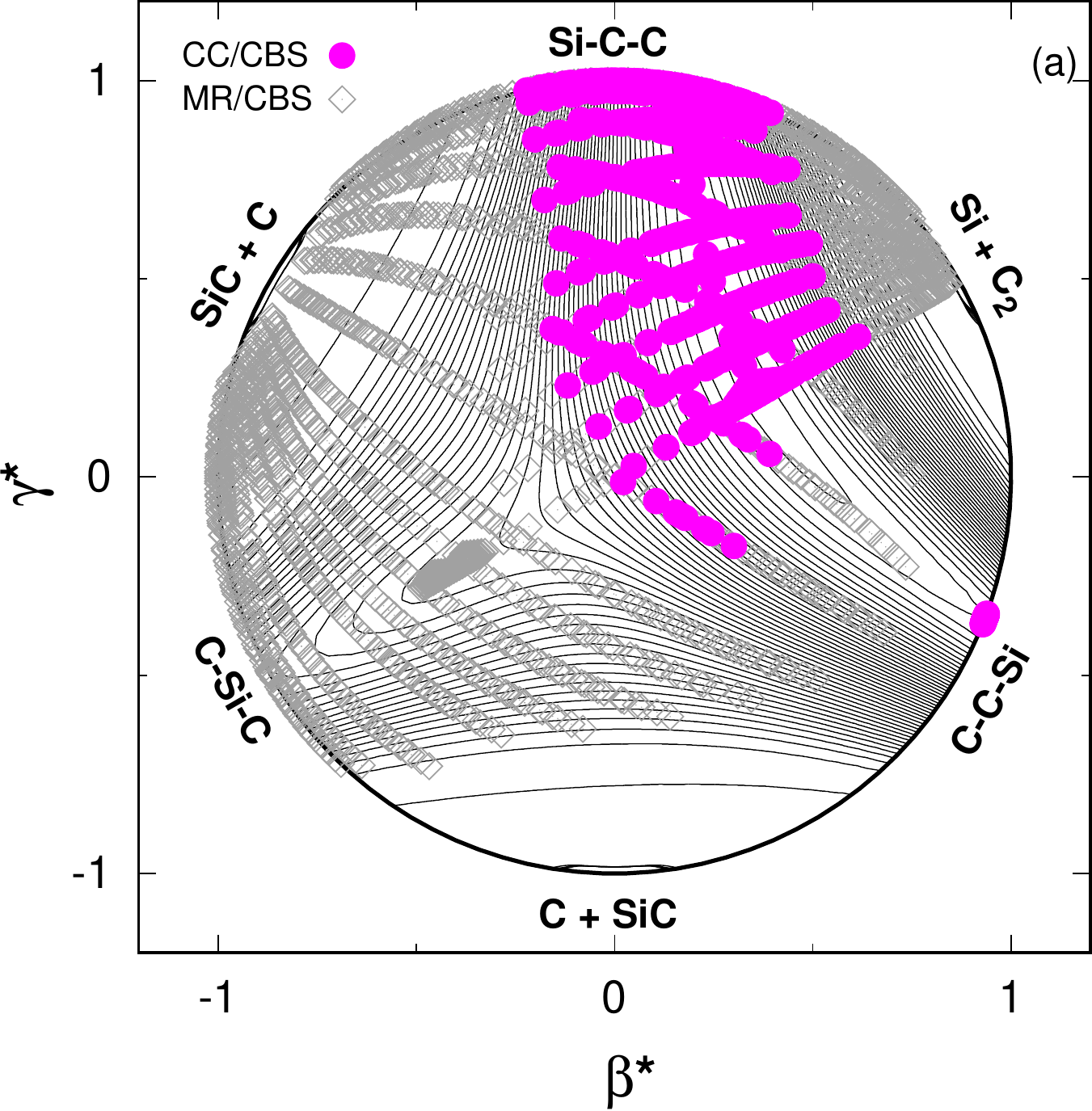}}}
\hfill
\subfloat{{\includegraphics[width=0.495\linewidth]{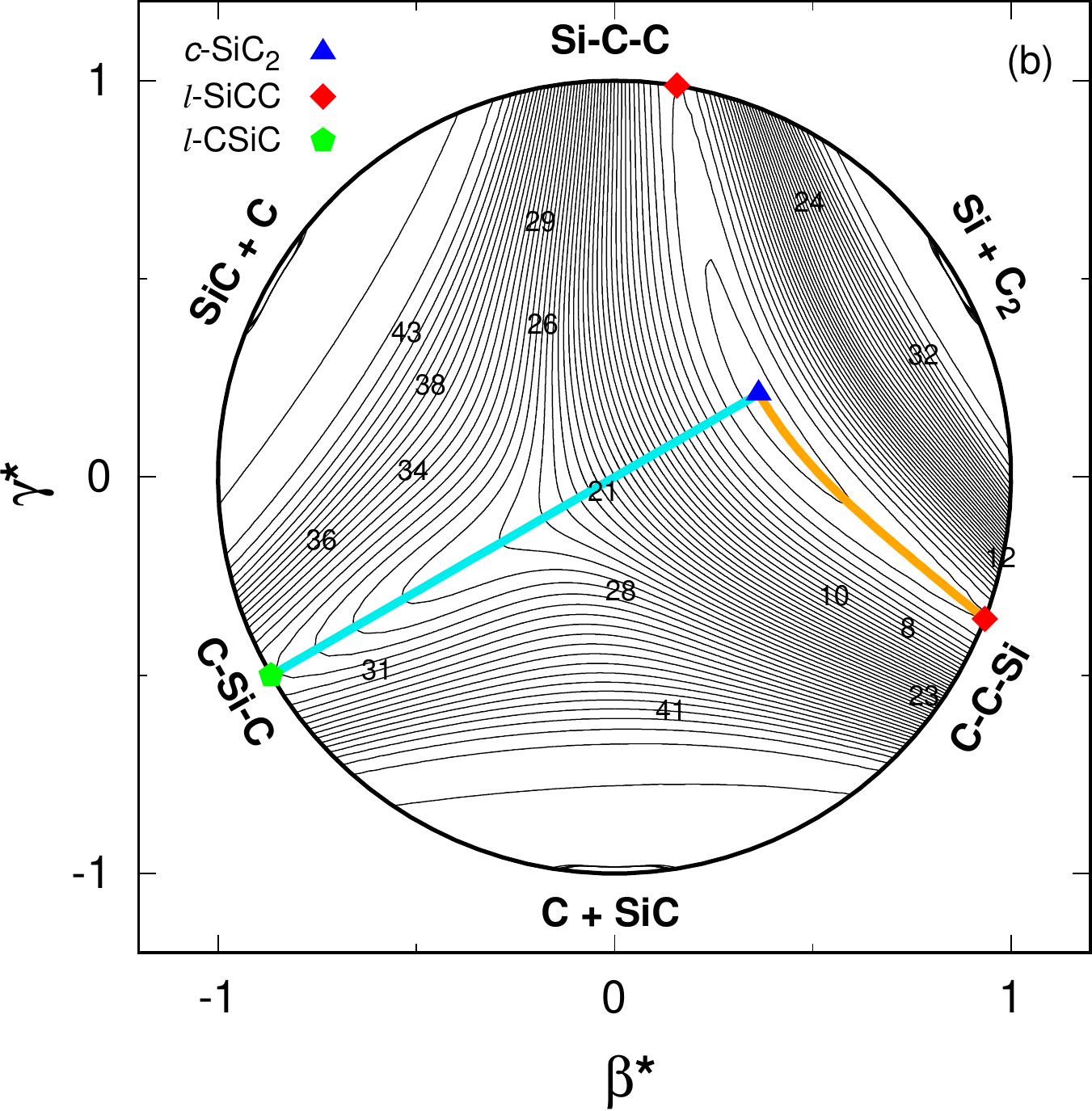}}}
\caption{\footnotesize Relaxed triangular plot in scaled hyperspherical coordinates~\citep{VAR87:455} [$\beta^{\star}$~and~$\gamma^{\star}$; {see Eq.~(\ref{eq:hypercoord})}] of the ground-state CHIPR PES of SiC$_{2}$ showing~(a).~the distribution of \emph{ab initio} CCSD(T)/CBS {(magenta solid circles)} and MRCI(Q)/CBS {(gray open diamonds)} calibration 
data set;~(b).~its global topographical attributes, location of stationary points (indicated by symbols), and isomerization pathways shown later in $1D$ in Figure~\ref{fig:pess}. {In both plots, linear geometries lie at the border of the physical circle, while
the $C_{2v}$ line connects $\mathrm{Si}\!+\!\mathrm{C_{2}}$ to $c$-$\mathrm{SiC}_{2}$ to $\ell$-$\mathrm{CSiC}$.  The origin ($\beta^{\star}\!=\!0$~and~$\gamma^{\star}\!=\!0$) 
defines a $D_{3h}$ configuration and $C_{s}$ structures are elsewhere. The location of all atom+diatom dissociation channels are properly indicated}. Contours are equally spaced 
by $0.007\,\rm E_{h}$ starting at $-0.5\,\rm E_{h}$. 
The zero of energy is set relative to the infinitely separated C+C+Si atoms. {The corresponding $3D$ version of plot~(b) is shown later in Figure~\ref{fig:traj}.}} 
\label{fig:hyper}
\end{figure*}
\begin{figure*}[htb!]
\begin{minipage}[c]{0.6\textwidth}
\includegraphics[angle=0,width=1\linewidth]{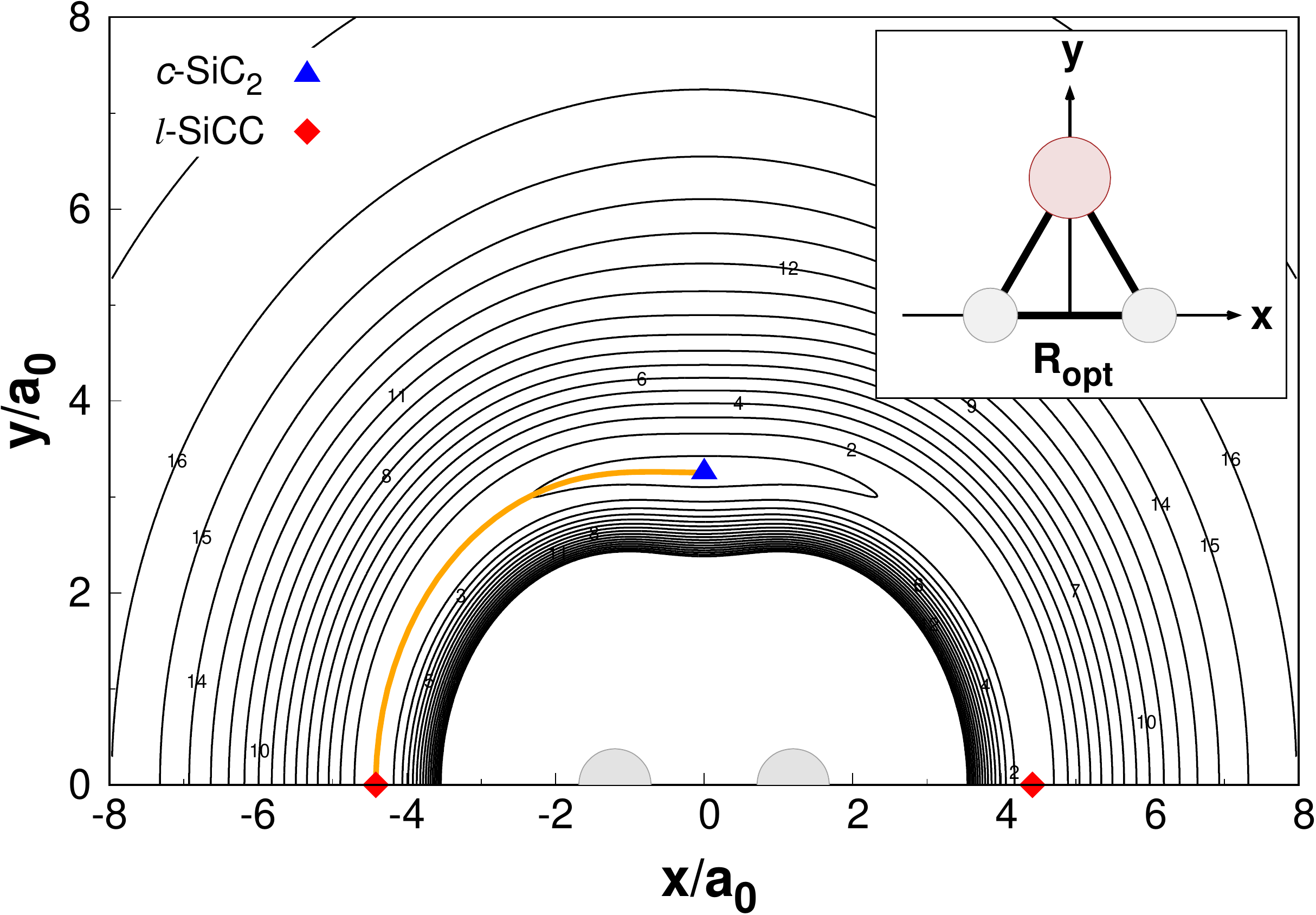}
\end{minipage}\hfill
\begin{minipage}[c]{0.4\textwidth}
\caption{\footnotesize CHIPR contour plot for a Si atom moving around a 
partially relaxed C$_{2}$ diatom (2.2\,$\leqslant$\,$r$/$a_{0}$\,$\leqslant$\,2.6), which lies along the x axis with the center of the bond fixed at the origin. { x and y coordinates give the position of Si
with respect to the origin. Linear geometries are defined by $x\!\neq\!0$ and $y\!=\!0$, while the $x\!=\!0$ and $y\!\neq\!0$ line describes $C_{2v}$ configurations; $C_{s}$ structures are elsewhere. 
} Contours are 
equally spaced by $0.015\,\rm E_{h}$ starting at $-0.5\,\rm E_{h}$. The zero of energy is set relative 
to the infinitely separated C+C+Si atoms. Solid color line represents the minimum energy path shown in $1D$ in Figure~\ref{fig:pess}(a).} 
\label{fig:siaroundc2}
\end{minipage}
\end{figure*}

In Eq.~(\ref{eq:pes}), the CHIPR diatomic curves 
are expressed by the general form~\cite{ROC021:107556}
\begin{equation}\label{eq:chiprdiat}
V^{(2)}(R)=\frac{Z_{\rm A}Z_{\rm B}}{R}\sum_{k=1}^{L}\!C_{k}\,y^{k}, 
\end{equation}
where $Z_{\rm A}$ and $Z_{\rm B}$ denote the nuclear 
charges of atoms A and B and the $C_{k}$'s are expansion coefficients; 
the $y$ coordinate is herein defined as a linear combination of $R$-dependent basis functions~\cite{ROC021:107556} (see Eq.~(\ref{eq:chiprcontract}) below). 
In turn, $V^{(3)}$ in Eq.~(\ref{eq:pes}) 
is represented via CHIPR's three-body model which for AB$_{2}$-type species 
assumes the simplified form~\cite{ROC021:107556,VAR013:134117,VAR013:408}
\begin{equation}\label{eq:chiprtriat}
V^{(3)}(\mathbf{R})=\!\!\!\sum_{i,j,k=0}^{L}\!C_{i,j,k}
\left[y_1^{i}(y_2^{j}y_3^{k}+y_2^{k}y_3^{j})\right].
\end{equation}
In the above equation, $C_{i,j,k}$ are expansion coefficients 
of a $L^{th}$-degree polynomial, 
and the $y_{p}$'s\,$(p\!=\!1,2,3)$ are (transformed) coordinates. 
These latter are expressed in terms of distributed-origin constracted basis sets~\cite{ROC021:107556} 
\begin{equation}\label{eq:chiprcontract}
y_p=\left(\sum_{\alpha=1}^{M-1}\!c_{\alpha}\,\phi^{[1]}_{p,\alpha}\right) + c_{M}\phi^{[2]}_{p,M},   
\end{equation}
where 
\begin{equation}\label{eq:chiprprimitive1}
\phi^{[1]}_{p,\alpha}={\rm sech}\left[\xi_{p,\alpha}\left(R_{p}\!-\!R_{p,\alpha}^{\mathrm{ref}}\right)\right],
\end{equation}
and
\begin{equation}\label{eq:chiprprimitive2}
\phi^{[2]}_{p,M}=\left[\frac{{\rm tanh}\left(\frac{1}{5}R_{p}\right)}
{R_{p}}\right]^{6}\hspace*{-0.2cm}{\rm sech}\left[\xi_{p,M}\left(R_{p}\!-\!R_{p,\alpha}^{\mathrm{ref}}\right)\right]     
\end{equation}
are primitive bases with origin at $R^{\mathrm{ref}}$ and the 
$\xi$'s are non-linear parameters. All steps involved in the calibration of Eqs.~(\ref{eq:chiprdiat})-(\ref{eq:chiprprimitive2}) using \emph{ab initio} data points  
are fully described in Refs.~\citenum{ROC020:106913}~and~\citenum{ROC021:107556}, with the reader being addressed 
to them for further details. 
Note that, to obtain the global analytic form of the PES [Eq.~(\ref{eq:pes})], we herein employ the newly-developed CHIPR-4.0 program~\cite{ROC021:107556}. With this code, the final CHIPR diatomic potentials of $\mathrm{C_{2}}$ and $\mathrm{SiC}$ [Eq.~(\ref{eq:chiprdiat})] were calibrated using MR/CBS points 
with rmsds of $1.1$~and~$0.3\,\mathrm{cm^{-1}}$, respectively. 
For completeness, they 
are plotted in Figure~S1. 
As for the three-body term, all 3682 \emph{ab initio} dual-level CC/MR CBS  
points could be least-squares 
fitted to Eq.~(\ref{eq:chiprtriat}) with chemical accuracy ($\mathrm{rmsd}\!=\!0.9\,\rm kcal\,mol^{-1}$). 
{ The weights ($W$) so employed were: $W\!=\!1$ for calculated points with 
energies $E\leq\!50\,\mathrm{kcal\,mol^{-1}}$ above the global $C_{2v}$ minimum, $W\!=\!0.7$ for those within the interval $50\!<\!E/\mathrm{kcal\,mol^{-1}}\!\leq\!135$, and $W\!=\!0.2$ for geometries with $E\!>\!135\,\mathrm{kcal\,mol^{-1}}$ above $c$-$\mathrm{SiC}_{2}$.}
Our fit involves a total of 
180 linear coefficients in the polynomial expansion [$L\!=\!11$ in Eq.~(\ref{eq:chiprtriat})]; see Tables~S1-S4 of the SM to access the 
numerical coefficients of all parameters resulting from the fit. 
Figure~S2 also portrays some representative cuts of the final analytic CHIPR potential [Eq.~(\ref{eq:pes})]  
alongside the corresponding \emph{ab initio} ones. 
Table~\ref{tab:rmsd} displays the stratified rmsd, 
while {{Figure~\ref{fig:devdist} shows the distribution of errors of the fitted data set}}. 
{ Accordingly, we note that $\sim\!82\%$ of the data is herein fitted with 
$0.9\,\rm kcal\,mol^{-1}$ accuracy. Moreover, Figure~\ref{fig:devdist} indicates that most of the calculated grid points  
(95\% of the total population) are primarily distributed within the 
$300\,\mathrm{kcal\,mol^{-1}}$ range above $c$-$\mathrm{SiC}_{2}$, thence approximately 
spanning the energy interval of up to its complete 
atomization~\cite{DEU94:387,OYE011:094103} (if we consider the atom+diatom 
geometries utilized to calibrate the diatomic curves). The high-energy points, particularly 
those within $300\!<\!E/\mathrm{kcal\,mol^{-1}}\!\leq\!1200$ (see Table~\ref{tab:rmsd} and Figure~\ref{fig:devdist}), are characterized by short 
CC and/or SiC bond distances which, despite carrying lower weights in the least-squares fitting procedure (see above), are shown to be important to properly model the repulsive 
walls of the global potential, preventing the three-body term [$V^{(3)}$ in Eq.~(\ref{eq:pes})] from attaining large negative values at these regions. 
}

In relation to our combined CC/MR protocol, we should mention that, 
{despite being extrapolated to the CBS limit}, 
the two \emph{ab initio} theories unavoidably diverge, {specially}     
at long distances; a prototypical case is illustrated in Figure~\ref{fig:cut}. 
This latter is clearly due to single-reference CC which is not expected to properly describe dissociation~\cite{SZA012:108}. These CC points, whenever present, were eliminated from the fit, warranting a smooth transition between the 
two data sets and the lowest rmsd; see Figures~\ref{fig:cut}~and~S2.  
{In the valence region, correlation energy differences between CC and MR also exist (even at CBS limit) but are less evident (Figure~\ref{fig:cut}), showing the smallest deviations near the global minimum; for example, at $c$-$\mathrm{SiC}_{2}$, the CC/CBS and MR/CBS total energies differ by $\sim\!71\,\mu\si{\hartree}$, a value that compares quite well with the corresponding estimate of $\sim\!11\,\mu\si{\hartree}$ calculated using   
CCSD(T)-F12b/VQZ-F12 and MRCI(Q)-F12/VQZ-F12 energies. These inherent discrepancies in CC and MR correlation energies are expected to increase when going up in energy, likely attaining larger values at long-range distances (Figure~\ref{fig:cut}). Note, however, that, because the low-energy part of our potential is primarily sampled by CC/CBS points [Figure~\ref{fig:hyper}(a)], we expect that the existence of such a CC/MR seam (lying higher in energy) influences little the final spectroscopic properties of the PES to be discussed next. 
} 
\begin{figure*}
\begin{minipage}[c]{0.6\textwidth}
\includegraphics[angle=0,width=1\linewidth]{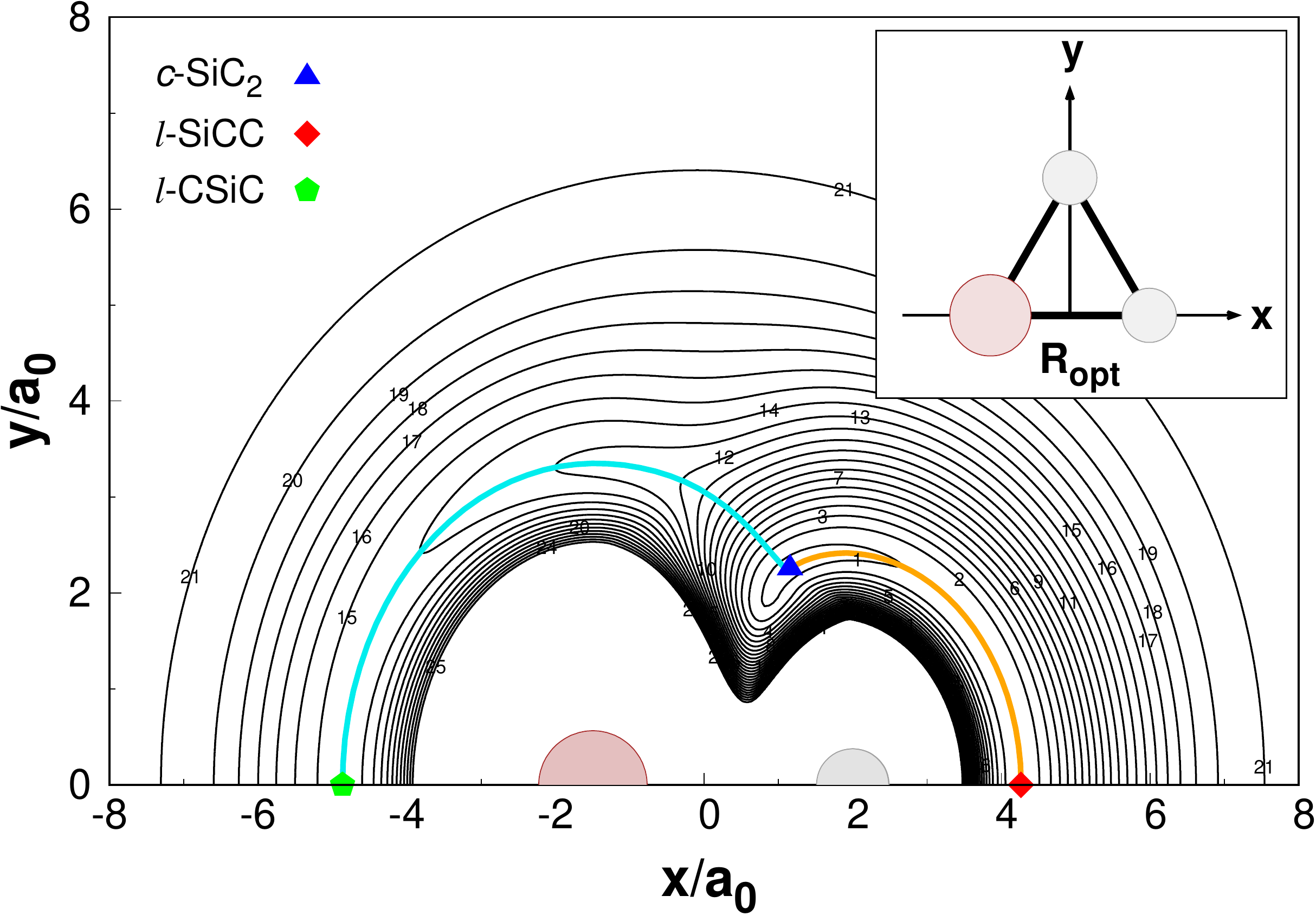}
\end{minipage}\hfill
\begin{minipage}[c]{0.4\textwidth}
\caption{\footnotesize CHIPR contour plot for a C atom moving around a 
partially relaxed SiC diatom (3.0\,$\leqslant$\,$r$/$a_{0}$\,$\leqslant$\,4.0), which lies along the x axis with origin at its center of mass. { x and y coordinates define the position of C 
with respect to the origin.} Contours are 
equally spaced by $0.015\,\rm E_{h}$ starting at $-0.5\,\rm E_{h}$. The zero of energy is set relative 
to the infinitely separated C+C+Si atoms. Solid color lines represent the minimum energy paths shown in $1D$ in Figure~\ref{fig:pess}(a)~and~(b).}
\label{fig:caroundsic}
\end{minipage}
\end{figure*}

\section{Features of PES}\label{sec:feat}
All major features of the final CHIPR PES are depicted in Figures~\ref{fig:hyper}-\ref{fig:traj}. 
The properties of its stationary points are collected in Table~\ref{tab:statpttab} wherein the most accurate results 
from the literature~\cite{FOR015:13,KOP016:2395,THA84:L45,MIC84:3556,BUT91:1,ROS94:4110} as
well as our own \emph{ab initio} CC and MR values 
are also included for comparison. {Note that, to allow for a complete visualization of all 
topographical attributes of our global CHIPR analytic potential, 
Figure~\ref{fig:hyper} shows a 
relaxed-triangular contour plot in 
scaled hyperspherical coordinates~\citep{VAR87:455},~$\beta^{\star}$=$\beta/Q$~and~$\gamma^{\star}$=$\gamma/Q$, where
\begin{equation} \label{eq:hypercoord}
\begin{pmatrix} Q \\ \beta \\ \gamma \end{pmatrix} = 
\begin{pmatrix} 1 & 1 & 1 \\ 0 & \sqrt{3} & -\sqrt{3} \\ 
2 & -1 & -1 \end{pmatrix} 
\begin{pmatrix} R_{1}^{2} \\ R_{2}^{2} \\ R_{3}^{2} \end{pmatrix},    
\end{equation}
and $R_{1}$,~$R_{2}$,~and~$R_{3}$ are interatomic distances.
Thus, by relaxing the ``size'' $Q$ of the molecule 
such as to give the lowest energy 
for a given ``shape`` ($\beta$~and~$\gamma$) of the triangle formed by the three atoms, the contour plot shown in 
Figure~\ref{fig:hyper} is then obtained;  
see legend therein and Refs.~\citenum{GAL009:14424}~and~\citenum{ROC015:074302} for further details. The corresponding $3D$ version of this plot is shown later in Figure~\ref{fig:traj}. In turn, Figures~\ref{fig:siaroundc2}~and~\ref{fig:caroundsic} illustrate the PES for the Si and C atoms moving around relaxed C$_{2}$ and SiC fragments, respectively. They also summarize in a comprehensive manner all predicted stationary structures 
from the analytic CHIPR PES to be discussed next.}

\subsection{Valence region \& spectroscopic calculations}\label{subsec:valence}
\begin{table*}[htb!]
\centering
\caption{\footnotesize Structural {equilibrium} parameters (in valence coordinates, $R_{i}$ in $\mathrm{a_0}$, $\alpha$ in degrees), 
harmonic ($\omega_{i}$) and fundamental ($\nu_{i}$) frequencies (in $\mathrm{cm^{-1}}$) of the stationary
points on the ground-state singlet PES of SiC$_{2}$. Relative energies ($\Delta E$)  
are in $\mathrm{kcal\,mol^{-1}}$ and given 
with respect to the $C_{2v}$ global minimum.}
\label{tab:statpttab}
\begin{ruledtabular}
\begin{threeparttable}
\begin{tabular}{cl@{}ddddddd
}
 \multirow{2}{*}{Structure} & \multirow{2}{*}{Method\tnote{a}} &
  {\multirow{2}{*}{$R_1$}}  & 
  {\multirow{2}{*}{$R_2$}}  & 
  {\multirow{2}{*}{$\alpha$}} &
  {\multirow{2}{*}{$\Delta E$\tnote{a}}} &
  {\omega_{1}} & 
  {\omega_{2}} &
  {\omega_{3}} \\[0.5ex] 
           &                 &
                             & 
                             & 
                             &
                             &
  {(\nu_{1})} & 
  {(\nu_{2})} &
  {(\nu_{3})} \\[0.5ex]   
\hline \\[-2.25ex]
\multirow{12}{*}{\centering \shortstack{\includegraphics[width=0.0525\textwidth]{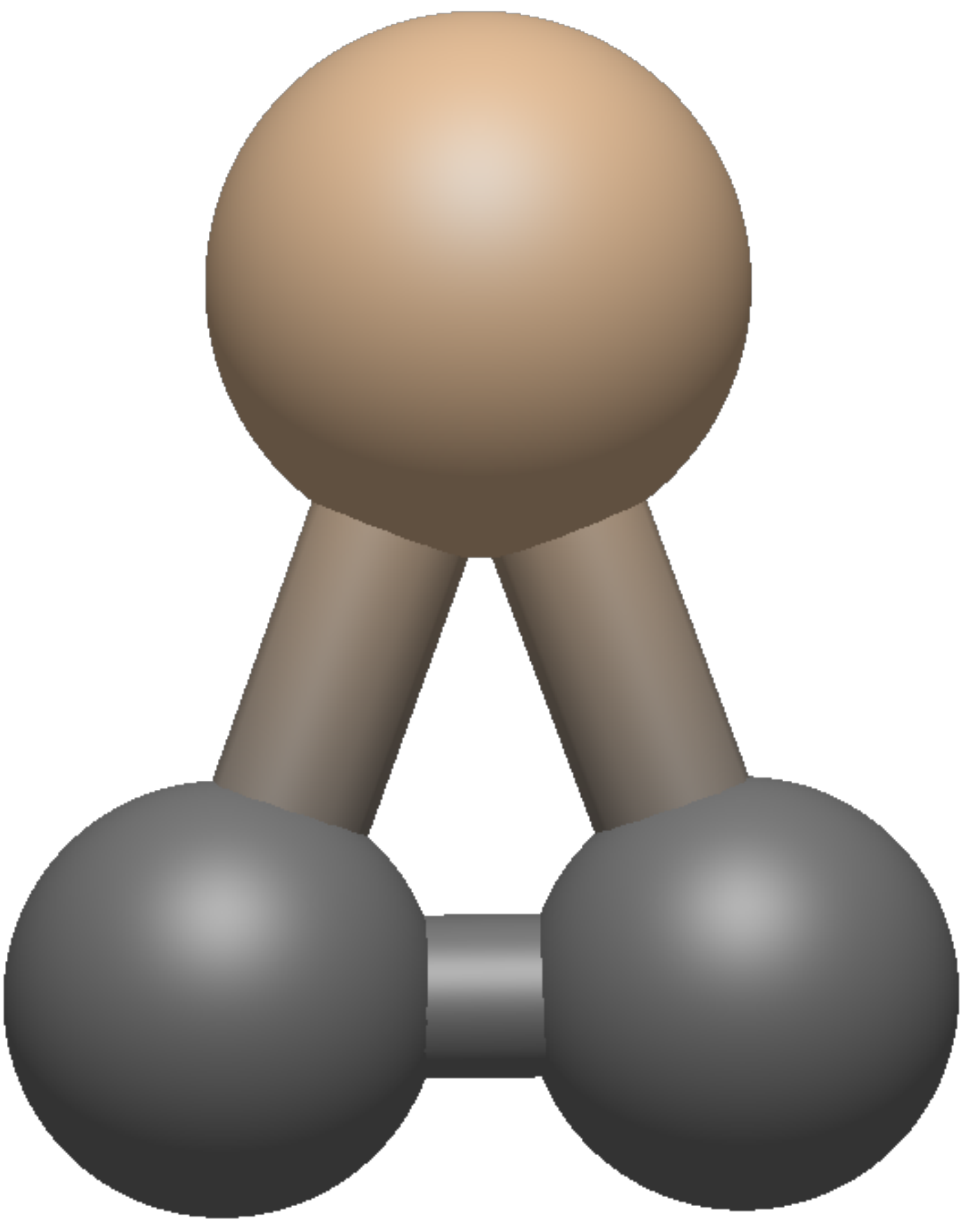}\\$c$-$\mathrm{SiC}_{2}$}} 
  & CC/AV$Q$Z                        & 3.478          & 3.478          &  40.5          &   0.0          &  1763.9           &   807.9                 &   180.5           \\
  & MR/AV$Q$Z                        & 3.488          & 3.488          &  40.6          &   0.0          &  1743.9           &   793.2                 &   223.2           \\
  & CC/CBS\tnote{b}                  &                &                &                &   0.0          &                   &                         &                   \\
  & MR/CBS\tnote{b}                  &                &                &                &   0.0          &                   &                         &                   \\
  & CcCR QFF\tnote{c}                & 3.460          & 3.460          &  40.5          &                &  1781.9           &   815.1                 &   201.4           \\
  &                                  &                &                &                &                & (1750.5)          & (844.7)                 &  (175.4)          \\
  & JK PES\tnote{d}                  & 3.460          & 3.460          &  40.6          &   0.0          &  1776.1           &   812.7                 &   214.6           \\
  &                                  &                &                &                &                & (1745.6)          & (837.9)                 &  (194.1)          \\
  & CHIPR PES                        & 3.468          & 3.468          &  40.5          &   0.0          &  1804.4           &   823.2                 &   201.6           \\
  &                                  &                &                &                &                & (1749.4)          & (840.1)                 &  (180.4)          \\
  & exp.                             & 3.459\tnote{e} & 3.459\tnote{e} &  40.6\tnote{e} &   0.0          &   1756.8\tnote{f} &   844.0\tnote{f}        &                   \\
  &                                  &                &                &                &                & (1746.0)\tnote{g} & (840.6)\tnote{g}        &  (196.4)\tnote{g} \\
\hline
\multirow{7}{*}{\centering \shortstack{\includegraphics[width=0.09\textwidth]{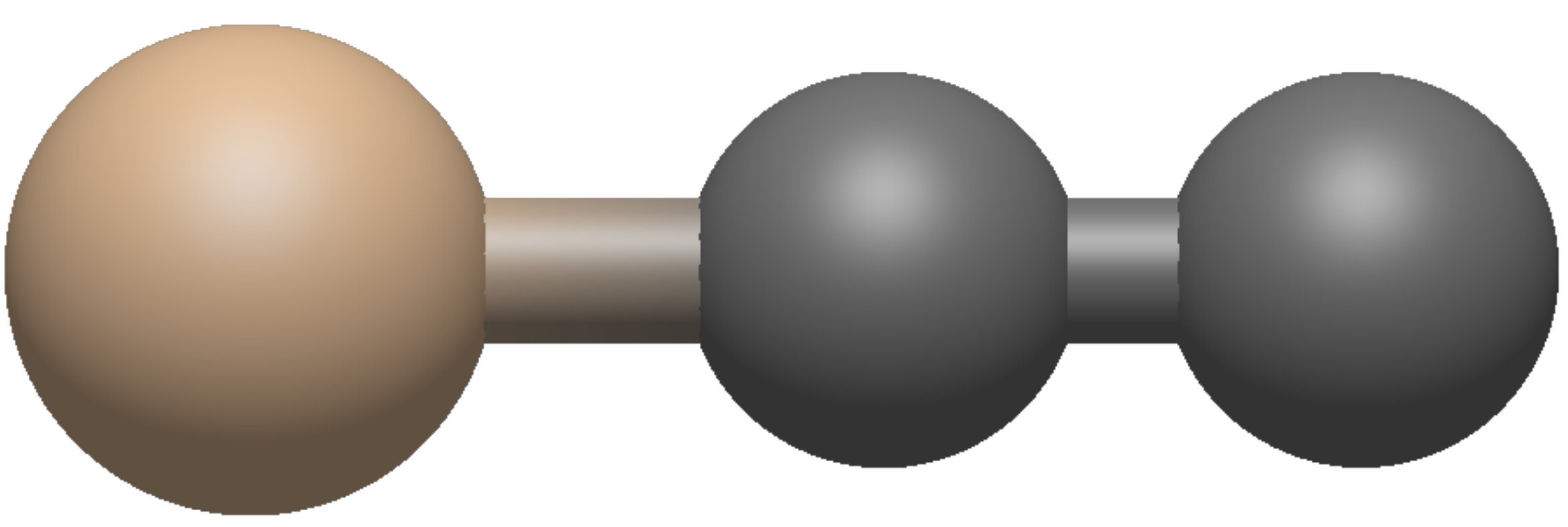}\\$\ell$-$\mathrm{SiCC}$}} 
  & CC/AV$Q$Z                        & 2.434          & 3.206          & 180.0          &   4.8          &     1887.9        &  787.4                  &   81.9{i}                \\
  & MR/AV$Q$Z                        & 2.456          & 3.231          & 180.0          &   4.2          &     1846.3        &  765.7                  &   55.7{i}                \\
  & CC/CBS\tnote{b}                  &                &                &                &   5.4          &                   &                         &                          \\
  & MR/CBS\tnote{b}                  &                &                &                &   4.5          &                   &                         &                          \\  
  & JK PES\tnote{d}                  & 2.425          & 3.192          & 180.0          &   5.1          &     1901.4        &  790.5                  &   82.6{i}                \\
  & CHIPR PES                        & 2.430          & 3.202          & 180.0          &   5.4          &     1893.6        &  783.5                  &  102.1{i}                \\
  & exp.                             &                &                &                &   5.4\pm0.6\tnote{h} &                   &                         &                          \\ 
\hline
\multirow{5}{*}{\centering \shortstack{\includegraphics[width=0.105\textwidth]{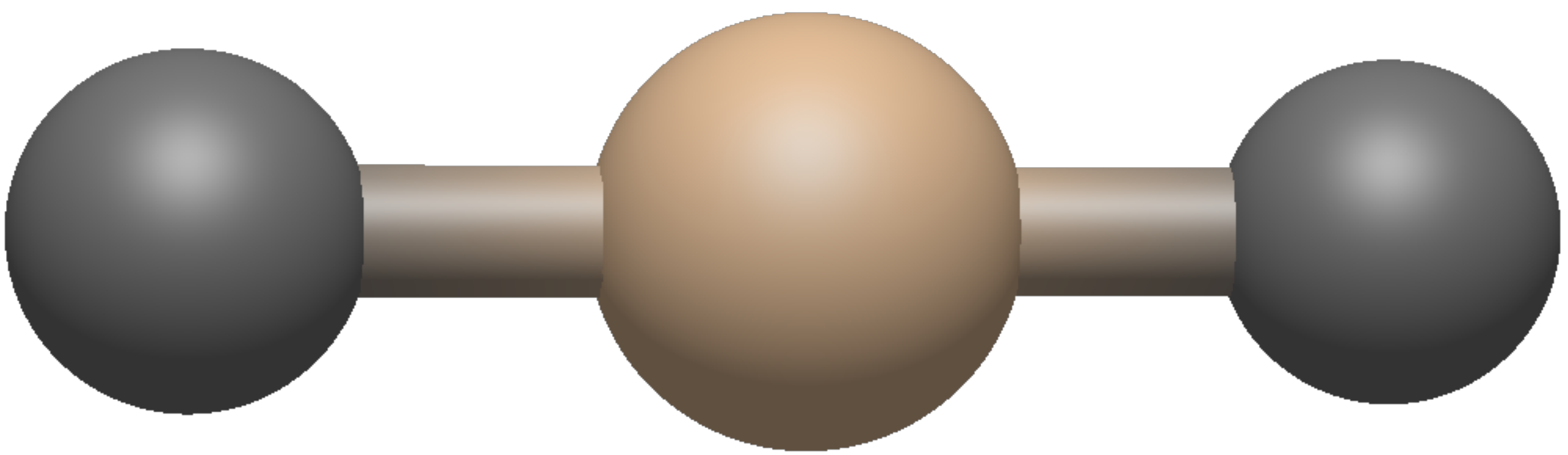}\\$\ell$-$\mathrm{CSiC}$}} 
  & CC/AV$Q$Z                        & 3.401          & 3.401          & 180.0          & 131.6          &      947.6        &  707.5                  &  177.8{i}               \\
  & MR/AV$Q$Z                        & 3.431          & 3.431          & 180.0          & 128.1          &      912.3        &  690.5                  &  162.6{i}               \\
  & CC/CBS\tnote{b}                  &                &                &                & 132.8          &                   &                         &                         \\
  & MR/CBS\tnote{b}                  &                &                &                & 129.2          &                   &                         &                         \\  
  & CHIPR PES                        & 3.409          & 3.409          & 180.0          & 129.2          &      902.8        &  729.2                  &  112.9{i}               \\  
\end{tabular}
\begin{tablenotes}[flushleft]
  \item[a]{{\footnotesize This work unless stated otherwise}.}
  \item[b]{{\footnotesize CC/CBS and MR/CBS single-point energies calculated at CHIPR PES stationary points}.}
  \item[c]{{\footnotesize Quartic force field of Ref.~\citenum{FOR015:13}}.}
  \item[d]{{\footnotesize Jacek Koput (JK) local PES of Ref.~\citenum{KOP016:2395}}.}
  \item[e]{{\footnotesize { Experimental equilibrium parameters reported in Ref.~\citenum{FOR015:13}. 
  The corresponding zero-point values are~\cite{BOG93:303} $R_{1,0}\!=\!R_{2,0}\!=\!3.463\mathrm{a_0}$ and $\alpha_{0}\!=\!40.505\,\si{\degree}$}}.}
  \item[f]{{\footnotesize { Experimental harmonic frequencies taken from Ref.~\citenum{BUT91:1}}}.} 
  \item[g]{{\footnotesize { Experimental fundamental frequencies taken from Ref.~\citenum{ROS94:4110}}}.}
  \item[g]{{\footnotesize { Potential energy barrier determined by Ross~\emph{et al.}~\cite{ROS94:4110} from experimental data}}.}
\end{tablenotes}
\end{threeparttable}
\end{ruledtabular}
\end{table*}
\begin{table}[htb!]
\centering
\caption{\footnotesize Calculated and observed vibrational term values (in $\mathrm{cm^{-1}}$) for $c$-$\mathrm{SiC}_{2}(^{1}A_1)$.}
\label{tab:vibcalc}
\begin{ruledtabular}
\begin{threeparttable}
\begin{tabular}{ccccddd
}
        &         &         &      &                 & \multicolumn{2}{c}{Calc} \\
\cline{6-7}        
$v_{1}$ & $v_{2}$ & $v_{3}$ & $\Gamma_{\mathrm{vib}}$ & {\rm Obs\tnote{a}} & {\rm CHIPR\tnote{b}} & {\rm JK\tnote{c}} \\
\hline
0       & 0       & 0       & $A_1$    &      0.0     &        0.0  &        0.0         \\
0       & 0       & 2       & $A_1$    &      352.85  &      326.1  &      349.6         \\
0       & 0       & 4       & $A_1$    &      605.33  &      573.5  &      600.3         \\
0       & 0       & 6       & $A_1$    &      814.87  &      793.4  &      809.2         \\
0       & 1       & 0       & $A_1$    &      840.6   &      840.1  &      837.9         \\
0       & 0       & 8       & $A_1$    &      1013.5  &      997.6  &     1005.1         \\
0       & 0       & 10      & $A_1$    &      1185.   &     1179.7  &     1179.2         \\
0       & 1       & 2       & $A_1$    &      1264.6  &     1239.0  &     1261.6         \\
0       & 0       & 12      & $A_1$    &      1350.48 &     1347.4  &     1342.6         \\
0       & 0       & 14      & $A_1$    &      1492.16 &     1490.4  &     1482.8         \\
0       & 1       & 4       & $A_1$    &      1556.7  &     1521.9  &     1549.0         \\
0       & 0       & 16      & $A_1$    &      1614.   &     1625.5  &     1609.3         \\
0       & 2       & 0       & $A_1$    &      1667.8  &     1667.7  &     1665.2         \\
1       & 0       & 0       & $A_1$    &      1746.0  &     1749.4  &     1745.6         \\
1       & 0       & 2       & $A_1$    &      2078.   &     2054.9  &     2075.1         \\
1       & 0       & 4       & $A_1$    &      2322.1  &     2297.4  &     2318.8         \\
0       & 3       & 0       & $A_1$    &      2465.7  &     2450.3  &     2460.4         \\
1       & 0       & 6       & $A_1$    &      2539.   &     2521.6  &     2530.7         \\
1       & 1       & 0       & $A_1$    &      2579.2  &     2588.8  &     2580.0         \\
1       & 0       & 8       & $A_1$    &      2735.   &     2733.6  &     2732.6         \\
1       & 0       & 10      & $A_1$    &      2918.   &     2924.9  &     2916.1         \\
0       & 4       & 0       & $A_1$    &      3303.   &     3292.4  &     3303.8         \\
1       & 2       & 0       & $A_1$    &      3406.6  &     3414.9  &     3405.6         \\
2       & 0       & 0       & $A_1$    &      3465.8  &     3464.2  &     3467.1         \\
2       & 1       & 0       & $A_1$    &      4299.   &     4304.2  &     4299.7         \\
2       & 2       & 0       & $A_1$    &      5122.   &     5125.0  &     5120.6         \\
3       & 0       & 0       & $A_1$    &      5164.   &     5155.5  &     5163.4         \\
0       & 0       & 1       & $B_1$    &      196.37  &      180.4  &      194.1         \\
0       & 0       & 3       & $B_1$    &      487.2   &      454.9  &      482.6         \\
0       & 0       & 5       & $B_1$    &      717.6   &      686.7  &      709.7         \\
0       & 1       & 1       & $B_1$    &      917.7   &      898.0  &      910.4         \\
0       & 0       & 7       & $B_1$    &      1072.2  &     1060.7  &     1070.1         \\
0       & 0       & 9       & $B_1$    &      1107.3  &     1094.8  &     1100.1         \\
0       & 1       & 3       & $B_1$    &      1271.   &     1266.1  &     1262.8         \\
0       & 0       & 11      & $B_1$    &      1412.   &     1385.5  &     1404.8         \\
0       & 0       & 13      & $B_1$    &      1436.5  &     1425.7  &     1429.7         \\
0       & 1       & 5       & $B_1$    &      1558.   &     1561.0  &     1550.2         \\
0       & 0       & 15      & $B_1$    &      1689.6  &     1686.1  &     1677.6         \\
0       & 1       & 7       & $B_1$    &      1883.   &     1879.9  &     1877.6         \\
1       & 0       & 1       & $B_1$    &      1925.   &     1917.5  &     1928.3         \\
0       & 2       & 1       & $B_1$    &      1955.   &     1941.3  &     1955.6         \\
1       & 0       & 3       & $B_1$    &      2201.   &     2179.3  &     2202.7         \\
1       & 0       & 5       & $B_1$    &      2430.   &     2421.7  &     2428.8         \\
1       & 0       & 7       & $B_1$    &      2627.   &     2628.8  &     2634.2         \\
\hline
\multicolumn{5}{c}{rmsd\tnote{d}}                     &       16.0  &        5.3         \\
\end{tabular}
\begin{tablenotes}[flushleft]
  \item[a]{{\footnotesize Experimental data from Ref.~\citenum{ROS94:4110}}.}
  \item[b]{{\footnotesize Calculated using CHIPR PES and {DVR3D}~\citep{TEN004:85}. {The zero point energy is $1400.1\,\mathrm{cm^{-1}}$.}}} 
  \item[c]{{\footnotesize Jacek Koput (JK) local PES. Data from Ref.~\citenum{KOP016:2395}.}}
  \item[d]{{\footnotesize Root-mean-square deviations with respect to experimenal data.}}
\end{tablenotes}
\end{threeparttable}
\end{ruledtabular}
\end{table}

\begin{figure}
\captionsetup[subfigure]{labelformat=empty}
\centering
\subfloat{{\includegraphics[width=1\linewidth]{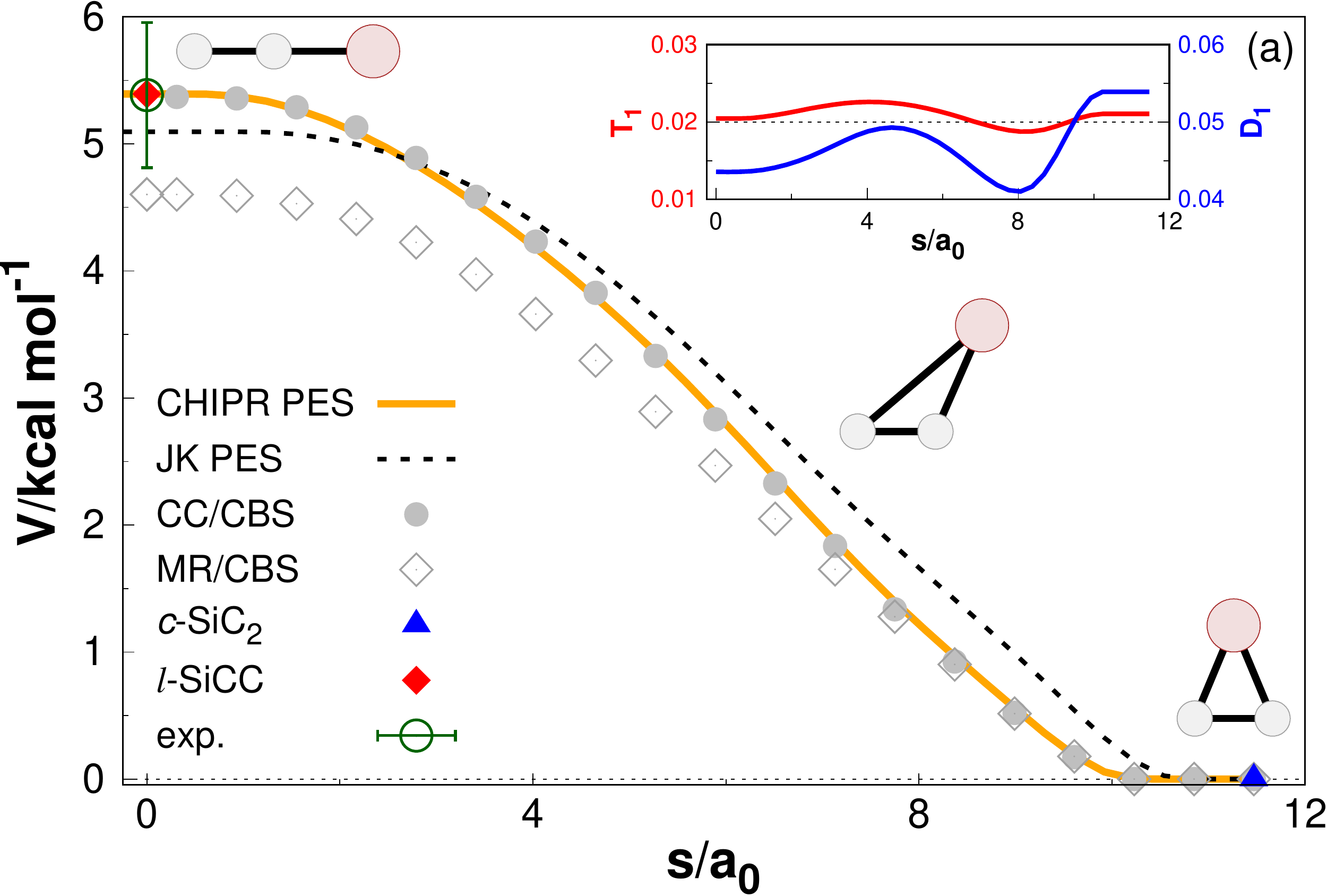}}}
\vfill 
\subfloat{{\includegraphics[width=1\linewidth]{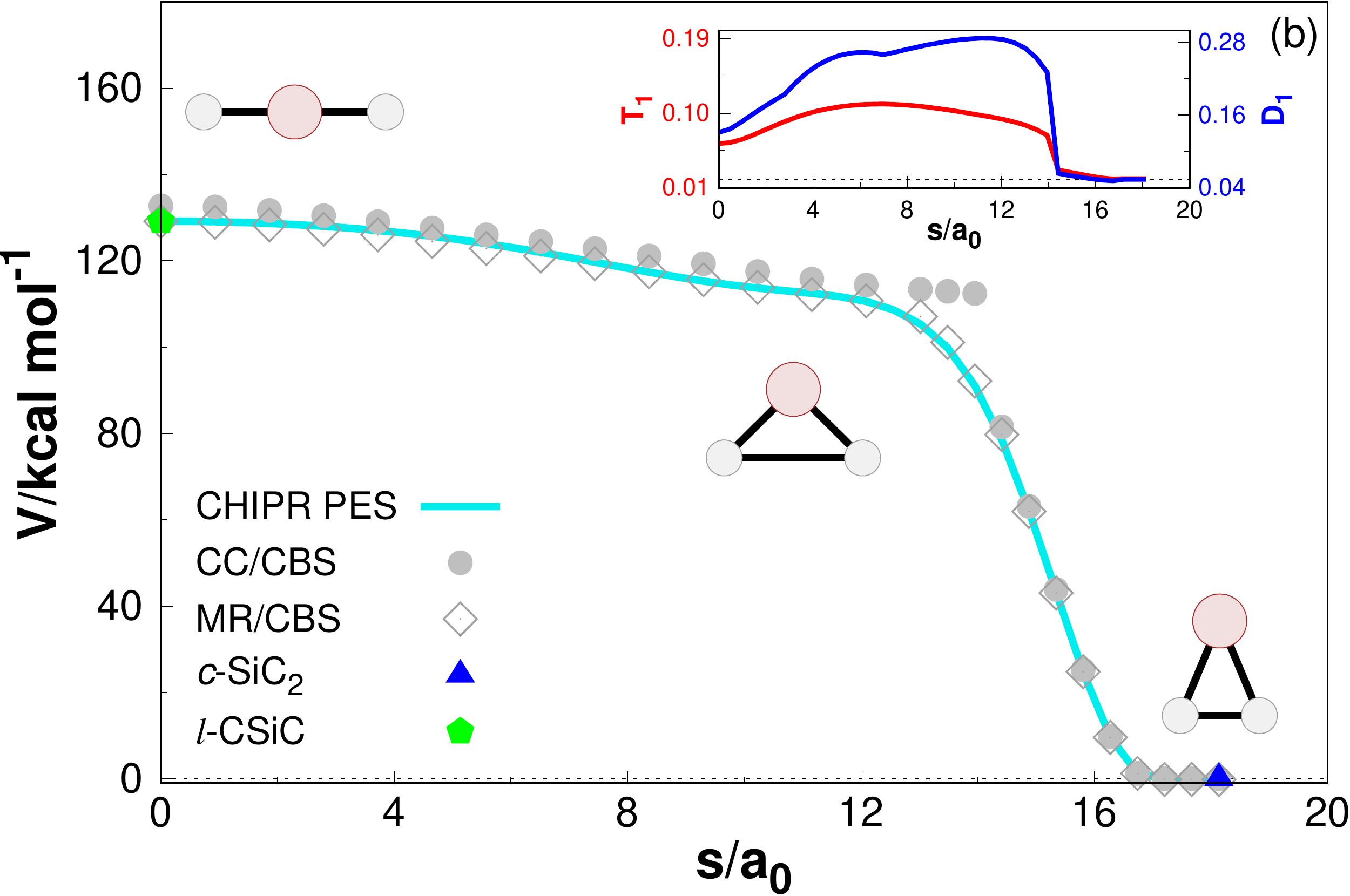}}}
\caption{\footnotesize CHIPR minimum energy paths ($s$ are reaction coordinates 
in mass-scaled a.u.) and potential energy barriers 
for the conversion of $c$-$\mathrm{SiC}_{2}$ to~(a).~$\ell$-SiCC~and~(b).~$\ell$-CSiC configurations. Solid circles  
and open diamonds indicate \emph{ab initio} CCSD(T)/CBS and MRCI(Q)/CBS single-point energies calculated at the predicted CHIPR geometries. In panel~(a)., the corresponding path obtained using    
the Jacek Koput (JK) local PES~\cite{KOP016:2395} {as well as the experimentally-derived
$c$-$\mathrm{SiC}_{2}$\,$\rightarrow$\,$\ell$-SiCC barrier reported by Ross~\emph{et al.}~\cite{ROS94:4110}} are also shown for comparison. {In panels~(a)~and~(b), MRCI(Q)/CBS and CCSD(T)/CBS points, respectively, were not included in the fit and are only plotted for comparison. The insets display the evolution of the coupled-cluster 
$T_{1}$ and $D_{1}$ diagnostics along the underlying paths as obtained from CCSD(T)/AV$5$Z calculations}.} 
\label{fig:pess}
\end{figure}

\begin{figure}
\captionsetup[subfigure]{labelformat=empty}
\centering
{\includegraphics[width=1\linewidth]{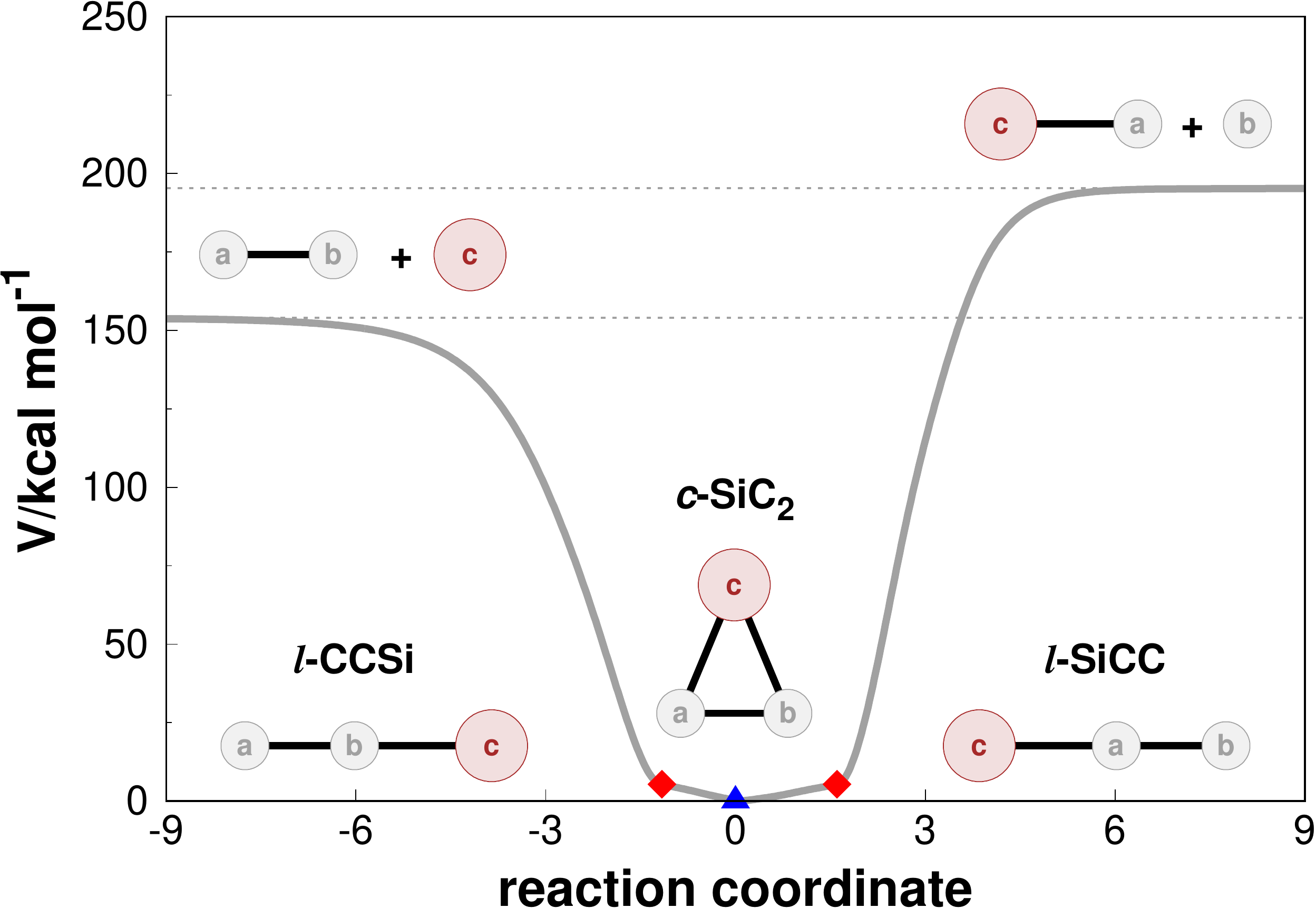}}
\caption{\footnotesize $1D$ cut of the CHIPR PES along 
the minimum energy path connecting C$_2$+Si and SiC+C 
via SiC$_2$ intermediates. Black dashed lines mark 
the associated energies of the infinitely separated atom+diatom fragments. 
The a,~b,~c labels are introduced to distinguish between symmetry-equivalent structures.} 
\label{fig:mep}
\end{figure}

According to Figures~\ref{fig:hyper}-\ref{fig:traj}, the predicted global mininum on the ground-state 
singlet PES corresponds to a cyclic $C_{2v}$ geometry, 
$c$-$\mathrm{SiC}_{2}(^{1}A_1)$. As Table~\ref{tab:statpttab} shows,  
its characteristic bond lengths and angle are $R_{1}(\mathrm{Si}\!-\!\mathrm{C})\!=\!R_{2}(\mathrm{Si}\!-\!\mathrm{C})\!=\!3.468\,\mathrm{a_{0}}$ and $\alpha(\angle\mathrm{C\!-\!Si\!-\!C})\!=\!40.5\si{\degree}$. These values are in excellent agreement with the most reliable theoretical 
estimates due to Fortenberry~\emph{et al.}~\cite{FOR015:13}~and~Koput~\cite{KOP016:2395}, differing by less than $0.008\,\mathrm{a_{0}}$/$0.1\si{\degree}$. Recall that these authors include, in addition to CBS energies, 
contributions from core-core/core-valence electron correlation and scalar relativistic effects in their local PESs; 
Koput~\cite{KOP016:2395} further accounts
for higher-order $\mathcal{N}$-particle electron correlation beyond CCSD(T). 
Close agreement is also found between 
the CHIPR's $c$-$\mathrm{SiC}_{2}$ data and experimental attributes taken from the literature~\cite{THA84:L45,MIC84:3556,BUT91:1,ROS94:4110}; see Table~\ref{tab:statpttab}. 
Indeed, our predicted $\mathrm{C}\!-\!\mathrm{C}$ ($\nu_{1}$) and $\mathrm{Si}\!-\!\mathrm{C}$ ($\nu_{2}$) stretching fundamentals 
reproduce exceedingly well ($\lesssim\!3.5\,\mathrm{cm^{-1}}$) the corresponding experimental values~\cite{ROS94:4110} {and are quite consistent with those calculated from the JK PES  (Table~\ref{tab:statpttab}).}
Yet, larger discrepancies (of up to $16\,\mathrm{cm^{-1}}$) are found for 
the large-amplitude $\nu_{3}$ fundamental associated with the internal 
rotation of the C$_2$ moiety. As noted elsewhere~\cite{NIE97:1195,KOP016:2395}, 
the proper description of the expectedly highly anharmonic potential along this mode [Figures~\ref{fig:siaroundc2}~and~\ref{fig:pess}(a)] 
requires an iterative treatment of the connected
triples ($T_{3}$) and quadruples ($T_{4}$) correlation contributions in the coupled-cluster expansion; 
this however would make the task of calculating the global PES of $\mathrm{SiC}_{2}$   
computationally unfeasible, even if limited to a smaller section of PES near $c$-$\mathrm{SiC}_{2}$ [Figure~\ref{fig:hyper}(a)]. Indeed, the corresponding $\nu_{3}$ value reported by Koput~\cite{KOP016:2395} differs by less than $3\,\mathrm{cm^{-1}}$ from its experimental estimate. Despite the expected lower performance of CHIPR relative to JK in predicting $\nu_{3}$, we note that our variationally-computed fundamentals for $c$-$\mathrm{SiC}_{2}$ still appear to be slightly more accurate than those reported using the CcCR QFF/VPT2 protocol~\cite{FOR015:13}, {even without considering here relativistic and core-valence correlation effects; see Table~\ref{tab:statpttab}.} 

To further assess the accuracy of the final CHIPR PES, we have carried out 
anharmonic vibrational calculations for higher excited modes using the {DVR3D} software suite~\citep{TEN004:85} 
and compared the results with the experimental term energies reported by Ross~\emph{et al.}~\cite{ROS94:4110}. All calculated data are gathered in Table~\ref{tab:vibcalc}. 
Also shown for comparison are the corresponding values reported from the JK local PES~\cite{KOP016:2395}.
Note that the vibrational band origins cover energies up 
to about $5200\,\mathrm{cm^{-1}}$ ($6600\,\mathrm{cm^{-1}}$) above 
the ground-state zero point level (bottom of the well) 
of $c$-$\mathrm{SiC}_{2}$ and excitations of up to as high as 16 quanta in $v_{3}$; the approximate
quantum numbers $v_{1}$ and $v_{2}$ refer to the 
$\mathrm{C}\!-\!\mathrm{C}$ and $\mathrm{Si}\!-\!\mathrm{C}$ stretching vibrations, while $v_{3}$ corresponds to the antisymmetric stretching of the triangular geometry.  
The results presented in Table~\ref{tab:vibcalc} show that 
our CHIPR PES reproduces remarkably well the vibrational spectrum of $c$-$\mathrm{SiC}_{2}$ with a rmsd of $16\,\mathrm{cm^{-1}}$ (as expected, the largest deviations are ascribed to overtones and combination bands involving $v_{3}$). This is quite astounding given the global, purely \emph{ab initio} nature of the PES and is clearly an asset of the present dual-level CC/MR 
protocol~\cite{GAL009:14424}. {It should be stressed that such a mixed protocol is herein 
devised to improve the spectroscopy of global potentials 
relative to global PESs calibrated solely using MR grid energies. 
Indeed, our experience shows (see, \emph{e.g.}, Refs.~\citenum{VAR002:1386}~and~\citenum{ROC018:36}) 
that purely MR global forms, despite accurately describing the bulk of the PESs, 
do in general a relatively poor job at reproducing experimental vibrational band origins of triatomics, showing rmsds of $\sim\!50\,\mathrm{cm^{-1}}$ or even greater. 
We reiterate that the lower 
performance of CHIPR when compared to the accurate JK local PES (see Table~\ref{tab:vibcalc}) 
is not surprising given the absence of higher-order effects in our CC calibration data, in addition to the fact that global analytic forms unavoidably entail larger fitting errors, even near the global minimum. In turn and differently from CHIPR, the JK potential cannot physically describe all dissociation channels and may show spurious features at regions of the PES characterized by 
large $\mathrm{C}\!-\!\mathrm{C}$ bond distances.} 
{Additionally, CHIPR describes by built-in the complete atomization of the system. 
Considering the $c$-$\mathrm{SiC}_{2}$'s anharmonic zero point energy ($1400.1\,\mathrm{cm^{-1}}$) and 
its stabilization energy relative to the C+C+Si atoms   
($-0.474269\,\rm E_{h}$), a total atomization energy of $293.6\,\mathrm{kcal\,mol^{-1}}$ 
is predicted from our PES. This value is in excellent agreement with the best theoretical estimate of $293.1\,\mathrm{kcal\,mol^{-1}}$ reported by Oyedepo~\emph{et al.}~\cite{OYE011:094103} using the MR-ccCA protocol~\cite{OYE011:094103} and the early G2 result by Deutsch~\emph{et al.}~\cite{DEU94:387} ($294.7\,\mathrm{kcal\,mol^{-1}}$); the last known experimental value is~\cite{DEU94:387} 
$301.0\pm7\,\mathrm{kcal\,mol^{-1}}$. 
}

As Figures~\ref{fig:hyper}(b)~and~\ref{fig:siaroundc2} portray, $c$-$\mathrm{SiC}_{2}$ is connected by two-symmetry equivalent linear ($C_{\infty v}$) transition states,  
$\ell$-$\mathrm{SiCC}(^{1}\Sigma^{+})$, located at $R_{1}(\mathrm{C}\!-\!\mathrm{C})\!=\!2.430\,\mathrm{a_{0}}$, 
$R_{2}(\mathrm{Si}\!-\!\mathrm{C})\!=\!3.202\,\mathrm{a_{0}}$ and $\alpha(\angle\mathrm{Si\!-\!C\!-\!C})\!=\!180.0\si{\degree}$ with an imaginary frequency of $102.1\,\mathrm{cm^{-1}}$. 
The corresponding minimum energy path (MEP) calculated~\cite{POLYRATE} from the PES is plotted in Figure~\ref{fig:pess}(a) and clearly represents the large-amplitude nearly-free pinwheel motion of C$_2$ around Si. Indeed, 
a close look at Figure~\ref{fig:pess}(a) shows that the CHIPR form accurately reproduces
the MEP at the CC/CBS level, with the corresponding MR/CBS path being actually lower in energy. 
{Suffice it to say that such MR/CBS points are only shown therein for comparison~--~they were not included in 
the fit as this region is only sampled by CC/CBS points (section~\ref{subsec:abinitioextrap}).}
Our best theoretical estimate (taken from the analytic PES) 
places $\ell$-$\mathrm{SiCC}$ at $5.4\,\mathrm{kcal\,mol^{-1}}$ ($1886.1\,\mathrm{cm^{-1}}$) 
above $c$-$\mathrm{SiC}_{2}$, in excellent agreement with the reported value  
of $5.1\,\mathrm{kcal\,mol^{-1}}$ ($1781.9\,\mathrm{cm^{-1}}$) by Koput~\cite{KOP016:2395}.  
Most notably, our predicted barrier to linearity is shown to match nearly
perfectly the corresponding experimental estimate  
of~\cite{ROS94:4110} $5.4\pm0.6\,\mathrm{kcal\,mol^{-1}}$ ($1883\pm200\,\mathrm{cm^{-1}}$). 
{These results 
provide compelling evidence that, at this level, CC appears to be more reliable in describing the $c$-$\mathrm{SiC}_{2}$/$\ell$-$\mathrm{SiCC}$ region, 
despite lying at the threshold of single-reference description with~\cite{LEE89:199,JAN98:423} $T_{1}\!\sim\!0.02$ and $D_{1}\!\sim\!0.05$; see inset of Figure~\ref{fig:pess}(a). 
The corresponding barrier predicted at MR/CC level is $\approx 0.9\,\mathrm{kcal\,mol^{-1}}$ lower than the CC/CBS estimate (Table~\ref{tab:statpttab}), being nearly coincident with  
the value of $4.5\,\mathrm{kcal\,mol^{-1}}$ reported by Koput at MR-ACPF/cc-pV6Z level~\cite{KOP016:2395}.} 

A notable aspect of the CHIPR PES, discussed previously in early studies~\cite{GRE83:895,ARU95:71,ZHA98:31}, is the existence of an auxiliary $D_{\infty h}$ transition state, 
$\ell$-$\mathrm{CSiC}(^{1}\Sigma^{+}_{g})$. 
As Table~\ref{tab:statpttab} shows, this linear form has characteristic bond lengths of $R_{1}(\mathrm{Si}\!-\!\mathrm{C})\!=\!R_{2}(\mathrm{Si}\!-\!\mathrm{C})\!=\!3.409\,\mathrm{a_{0}}$ and an imaginary frequency of $112.9\,\mathrm{cm^{-1}}$ along the bending coordinate. Its connection to $c$-$\mathrm{SiC}_{2}$ is perhaps 
best seen from the contour plots in Figures~\ref{fig:hyper}(b)~and~\ref{fig:caroundsic}; see the cyan solid lines represented therein. The associated isomerization pathway~\cite{POLYRATE} in $1D$ is presented in Figure~\ref{fig:pess}(b), wherein the major topographical valence attributes of 
the 
CHIPR PES across $C_{2v}$ geometries can be assessed. Accordingly, $\ell$-$\mathrm{CSiC}$ 
is predicted from our final CHIPR form to lie $129.2\,\mathrm{kcal\,mol^{-1}}$ above $c$-$\mathrm{SiC}_{2}$. 
Differently from $\ell$-$\mathrm{SiCC}$ which span a low energy region of the PES primarily 
sampled by CC/CBS points [Figures~\ref{fig:hyper}(a)], the description of $\ell$-$\mathrm{CSiC}$ and vicinities can only be accurately  
done at MR/CBS level. In fact, {as the inset of Figure~\ref{fig:pess}(b) shows}, at this region of the nuclear configuration space 
the predicted CC diagnostics [\emph{e.g.}, $T_{1}\!\approx\!0.11$ and $D_{1}\!\approx\!0.27$ 
halfway through the MEP] far exceed the accepted limiting values:  
$T_{1}\!\lesssim\!0.02$~\cite{LEE89:199}, $D_{1}\!\lesssim\!0.05$~\cite{JAN98:423}, thus clearly entailing a multi-reference approach. {This is explained by the presence of several 
low-lying excited electronic states in this region, as Figure~S3(a) illustrates.} Indeed, Figure~\ref{fig:pess}(b) evinces 
that the CHIPR form mimics excellently well the \emph{ab initio} MR/CBS data, with the predicted barrier to linearity matching exactly 
the one calculated at this level (Table~\ref{tab:statpttab}). {We further note that, 
in Figure~\ref{fig:pess}(b), 
the CC/CBS data shown  
are only plotted for comparison; they were not included in the calibration data set as 
this high-energy valence region of the PES (with $s\!\lesssim\!15\,\mathrm{a_{0}}$)
is sampled solely by MR/CBS calculations.}   

\subsection{Proof of concept: long-range region \& reaction dynamics calculations}\label{subsec:longrange}
\begin{figure}
\captionsetup[subfigure]{labelformat=empty}
\centering
\subfloat{{\includegraphics[width=1\linewidth]{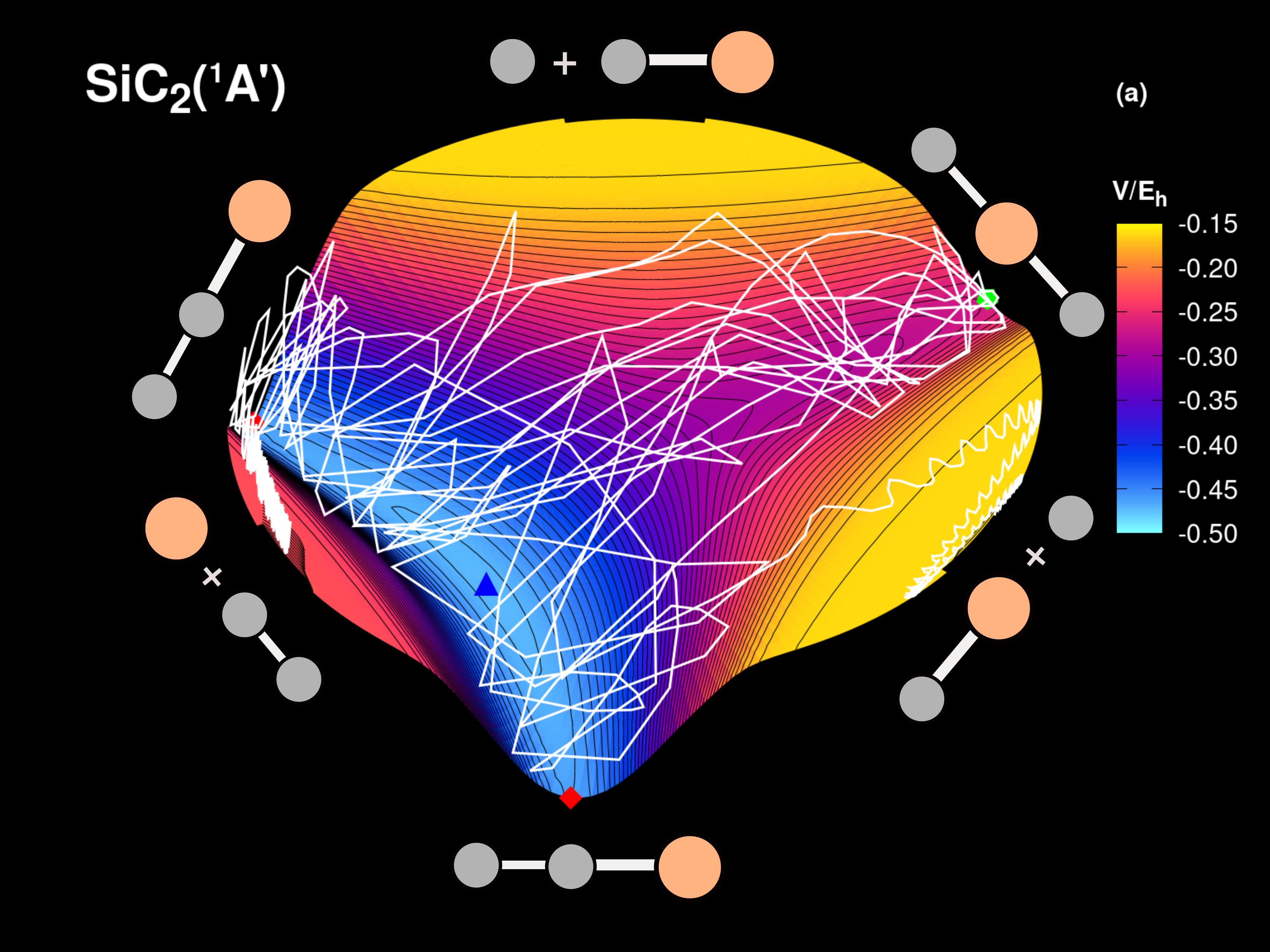}}}
\vfill 
\subfloat{{\includegraphics[width=1\linewidth]{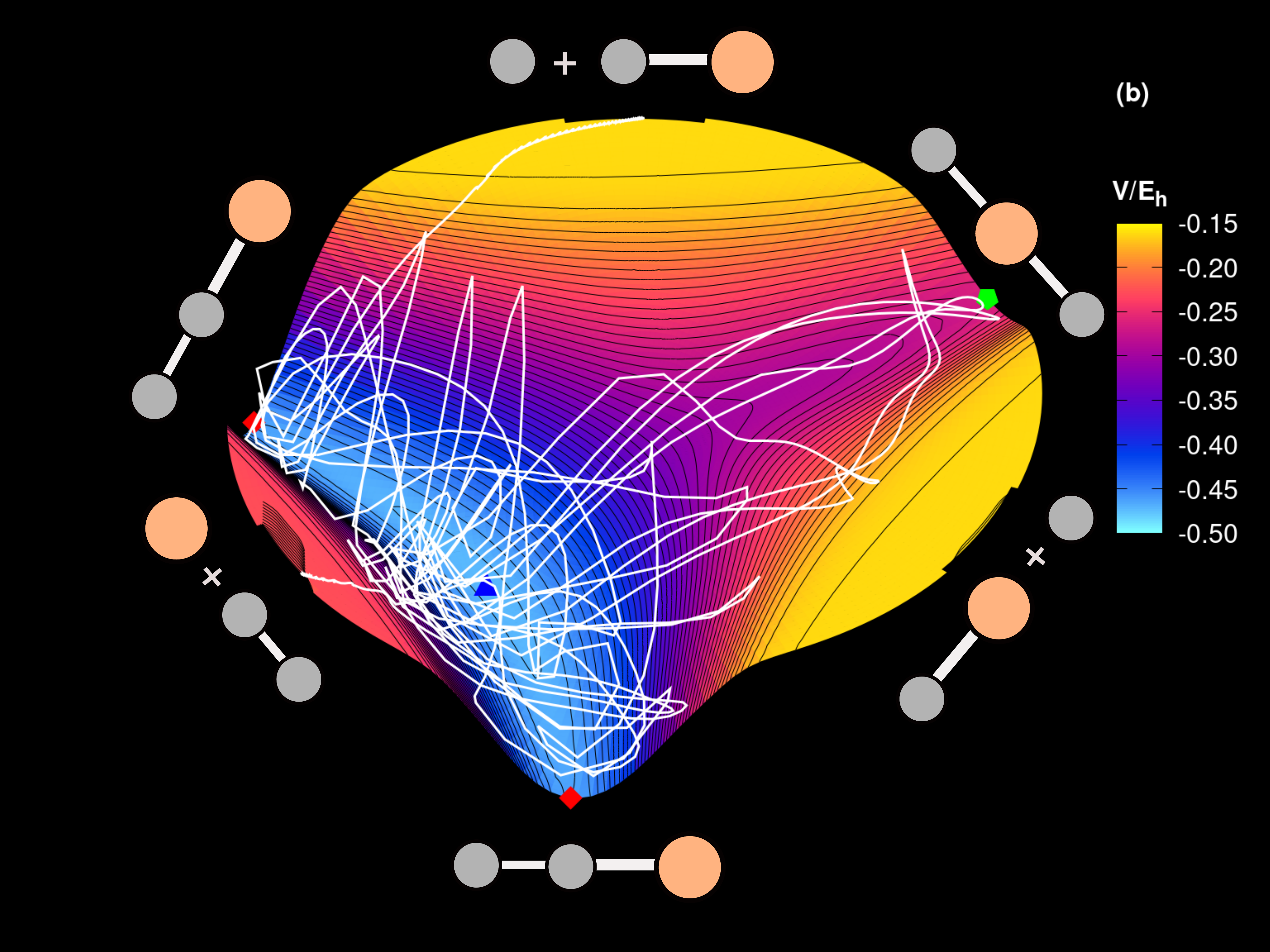}}}
\caption{\footnotesize Relaxed 3$D$ hyperspherical plots [see also Figure~\ref{fig:hyper}(b)] 
of the CHIPR PES of $\mathrm{SiC}_{2}(^{1}A')$ showing the time evolution (in coordinate space) of sample reactive quasi-classical trajectories (solid white lines) calculated using the {VENUS96C} code~\citep{VENUS} for $\mathrm{Si}(^{3}P)\!+\!\mathrm{C_{2}}(a^{3}\Pi_{u})\!\rightarrow\!\mathrm{SiC}(X^{3}\Pi)\!+\!\mathrm{C}(^{3}P)$ with distinct initial conditions~(a).~vibrationally excited 
C$_2(v\!=\!11)$ 
and collision energy of $1.0\,\mathrm{kcal\,mol^{-1}}$~and~(b).~ground-state C$_2$ 
and collision energy of $42.0\,\mathrm{kcal\,mol^{-1}}$. Stationary points and coordinates as in Figure~\ref{fig:hyper}(b). The zero of energy is set relative to the infinitely separated C+C+Si atoms. 
} 
\label{fig:traj}
\end{figure}
\begin{figure}[htb!]
\includegraphics[width=1\linewidth]{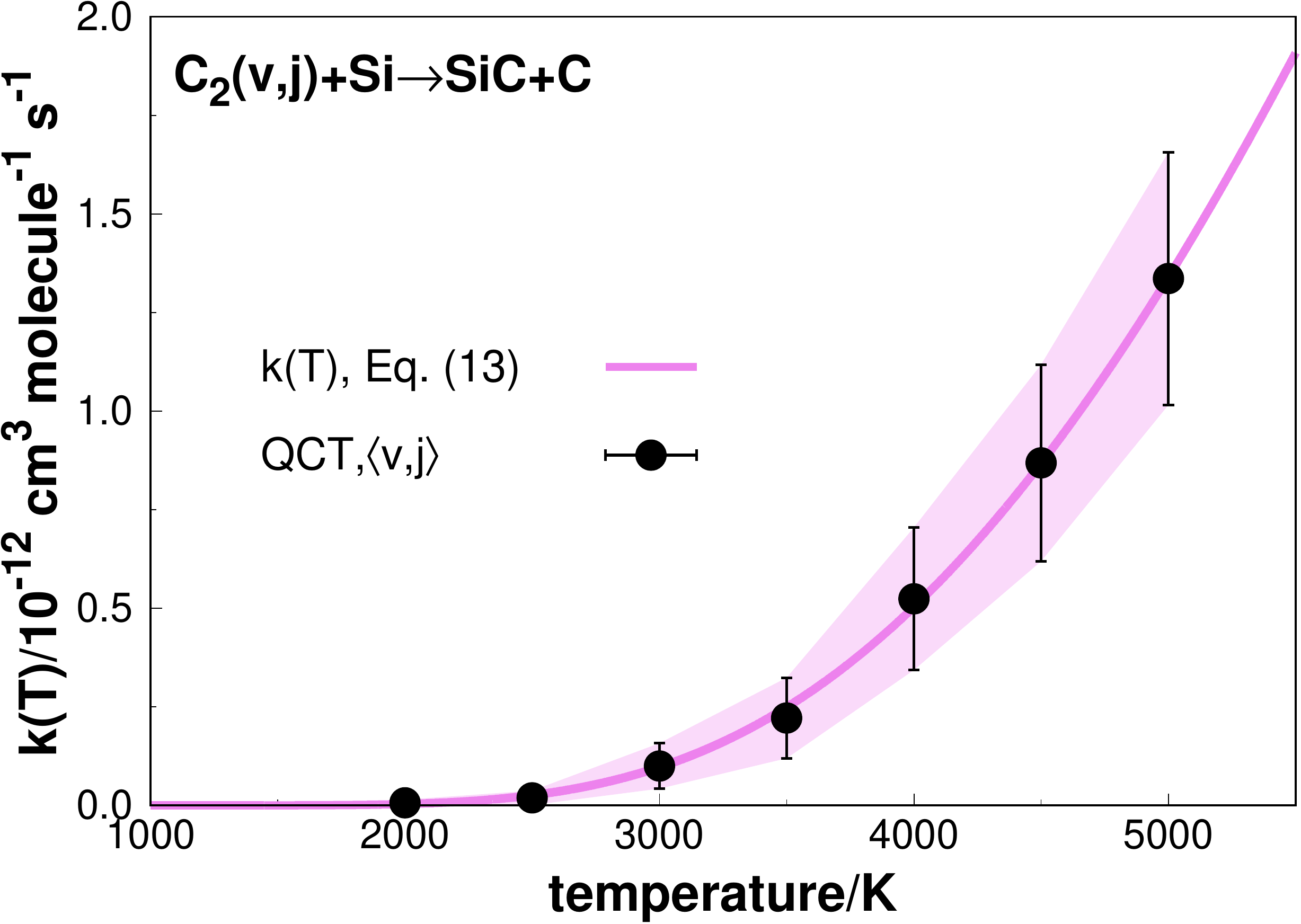}
\caption{\footnotesize Calculated rate constants and associated error bars for 
the $\mathrm{Si}(^{3}P)\!+\!\mathrm{C_{2}}(a^{3}\Pi_{u})\!\rightarrow\!\mathrm{SiC}(X^{3}\Pi)\!+\!\mathrm{C}(^{3}P)$ 
reaction within the temperature range of $2000\!\leq\!T/\si{\kelvin}\!\leq\!5000$. 
The lines show the predicted QCT thermally averaged results from Eq.~(\ref{eq:ratehigh}). For clarity, the QCT results are shown with 99.6\% ($3\sigma$) error bars.} 
\label{fig:rate}
\end{figure}
Apart from accurately modeling the valence (strongly-bound) chemical space,  
the contour plots shown in Figures~\ref{fig:hyper}-\ref{fig:caroundsic} evidently pinpoint the reliability of the CHIPR form to 
describe 
long-range and 
dissociation features of the $\mathrm{SiC}_{2}$ PES, in addition to naturally reflect its correct permutational symmetry. 
This is clearly an asset of the CHIPR~\cite{VAR013:054120,VAR013:408,ROC020:106913,ROC021:107556} formalism [namely, Eq.~(\ref{eq:chiprtriat})] and is the major deliverable of the present work. Figure~\ref{fig:mep} shows the calculated 
MEP for the chemical conversion of C$_2$+Si to SiC+C 
that 
proceeds  
via 
SiC$_2$ intermediates. Accordingly, both forward and reverse collision processes evolve without activation barriers for collinear atom-diatom
approaches, leading directly to the formation of $\ell$-$\mathrm{SiCC}$. 
This structure is subsequently converted to $c$-$\mathrm{SiC}_{2}$ by way of  
low-energy (nearly-free) C$_2$ internal rotations [Figure~\ref{fig:pess}(a)];
the stabilization energy of the $c$-$\mathrm{SiC}_{2}$ complex 
is predicted to be \emph{ca.}~$-154.1$~and~$-195.4\,\mathrm{kcal\,mol^{-1}}$ relative to the infinitely separated C$_2$+Si and SiC+C fragments, respectively, this former being quite close to the 
value of $-152.9\,\mathrm{kcal\,mol^{-1}}$ 
reported by Nielsen~\emph{et al.}~\cite{NIE97:1195} based on  
high-level focal point thermochemical analyses. 
Indeed, as Figure~\ref{fig:mep} shows, the $\mathrm{C_{2}\!+\!Si}\rightarrow\mathrm{SiC\!+\!C}$ reaction is highly endothermic 
($40.4\,\mathrm{kcal\,mol^{-1}}$, including the zero-point energies of the reactants and products) 
which makes this process feasible only in high-temperature environments, \emph{e.g.}, in the inner envelopes surrounding (late-type)  
carbon-rich 
stars~\cite{MAS018:A29,MCC019:7}, thence conceivably playing therein  
a key role in  
the formation of gas-phase SiC, and consequently solid SiC dust.  
Initial assessments indicate that, in order to effectively 
initiate such a reaction, $\mathrm{C_{2}}$ must 
be initially pumped~\cite{ZAN013:125} to higher vibrational states 
(up to at least $v\!=\!10$-11) or collide 
with a high-energy Si atom, with relative translational energies of the order of $\sim\!41\,\mathrm{kcal\,mol^{-1}}$ or higher; see, \emph{e.g.}, Figure~\ref{fig:traj}. 
These conditions can be 
fulfilled within the inner layers of the circumstellar shells      
of evolved C-stars (\emph{e.g.}, IRC+10216) 
characterized by temperatures of $\sim$1000-3000\,K or higher and where $\mathrm{C_{2}}(X^{1}\Sigma_{g}^{+},a^{3}\Pi_{u})$ and other silicon-carbon species are known to be particularly conspicuous~\cite{MCC019:7}. 
To further assess the reliability of such a reaction, we have run 
preliminary quasi-classical trajectory (QCT) calculations~\citep{PES99:171,VENUS} on the CHIPR PES using a locally modified 
version of the {VENUS96C} code~\citep{VENUS}; for a thorough description of the methodology here utilized, see Ref.~\citenum{ROC021:A142} and Table~S5. At the high temperature regime considered ($2000\!\leq\!T/\si{\kelvin}\!\leq\!5000$), the calculated thermal rate coefficients  
for $\mathrm{C_{2}\!+\!Si}\rightarrow\mathrm{SiC\!+\!C}$  
can be accurately represented by the Arrhenius-Kooij formula~\citep{LAI84:494}:
\begin{equation}\label{eq:ratehigh}
k(T)=A\left(\frac{T}{300}\right)^{B}\exp{\left(\frac{-C}{T}\right)},
\end{equation}
where $A\!=\!\SI{1.22582e-10}{}\,\rm cm^{3}\,molecule^{-1}\,s^{-1}$, $B\!=\!-0.161897$, and $C=\!20305.15\,\rm K$. This is 
plotted in Figure~\ref{fig:rate},  
together with the calculated QCT data which are numerically defined in Table~S5. 
Accordingly, the theoretically-predicted rate constants for $\mathrm{C_{2}\!+\!Si}\rightarrow\mathrm{SiC\!+\!C}$ show a positive 
temperature dependence, increasing steeply from $T\!=\!3000$\,K. 
This provides compelling evidence for its relevance in the gas-phase synthesis of SiC and related solid SiC dust formation in the innermost envelopes of C-stars~\cite{CER89:L25}. 
Further investigations in this direction are in order and a detailed account of the overall $\mathrm{C_{2}\!+\!Si}\rightarrow\mathrm{SiC\!+\!C}$ dynamics and kinetics 
undoubtedly requires a careful assessment of the possible contributions of other excited-states PESs {correlating to the same reactant/product channels}; 
this is clearly beyond the scope of our present preliminary analysis and will be the focus of future studies. Also of relevance is the reverse (barrierless and exothermic) $\mathrm{SiC\!+\!C}\rightarrow\mathrm{C_{2}\!+\!Si}$ reaction (Figure~\ref{fig:mep}) which, differently from $\mathrm{C_{2}\!+\!Si}\rightarrow\mathrm{SiC\!+\!C}$, surely occurs at cold and ultracold temperatures, hence dominated by long-range forces; indeed, the expected high reactivity of $\mathrm{SiC}$ with~\cite{MAC96:62} atomic C and O at low $T$s may help explain the lack of $\mathrm{SiC}$ detections in cold interstellar environments~\cite{HER89:205,WHI90:427,CHE021:5231}.

\section{Conclusions}\label{sec:conclusions}
We report the first global PES for ground-state $\mathrm{SiC}_{2}(^{1}A')$ based on CBS extrapolated \emph{ab initio} energies 
and the CHIPR method for the analytical modeling. 
By relying on a mixed CCSD(T) and MRCI(Q) protocol, we ensure 
that the final potential recovers much of the spectroscopy of its cyclic 
global minimum, while still permitting an accurate 
description of isomerization and fragmentation processes, all with the correct permutational symmetry as naturally warranted by CHIPR. 
Bound-state calculations performed anew have shown that the present purely-\emph{ab initio} CHIPR PES is capable of reproducing the experimental vibrational spectrum of cyclic $\mathrm{SiC}_{2}$ with a rmsd of $16\,\mathrm{cm^{-1}}$. 
{Despite not outperforming the spectroscopic quality of the most accurate local PES to date~\cite{KOP016:2395}, our proposed dual-level CCSD(T)/MRCI(Q) CBS protocol 
is expected to improve the spectroscopy of global ground-state PESs when compared with 
purely MRCI(Q)-based global forms.}
Further improvements can be so envisaged 
by either fine-tuning the theoretically-predicted potential parameters with input experimental information~\cite{VAR002:1386,VAR006:485} or morphing 
the original global form with a spectroscopically-accurate local potential~\cite{ROC018:36}. Aside from anharmonic vibrational calculations, 
the global nature of our CHIPR PES 
is further exploited by performing preliminary quasi-classical trajectory
calculations for the $\mathrm{C_{2}\!+\!Si}\rightarrow\mathrm{SiC\!+\!C}$ endothermic reaction. The calculated thermal rate
coefficients within the temperature range of $2000\!\leq\!T/\si{\kelvin}\!\leq\!5000$ hint for its prominence in the gas-phase synthesis of SiC and,  
presumably, SiC dust formation in the inner envelopes 
surrounding carbon-rich 
stars~\cite{MAS018:A29,MCC019:7}. 

\section*{Supplementary Material}
See the supplementary material to assess the performance of the PES alongside \emph{ab initio} grid data, the numerical coefficients of the final CHIPR analytic form as well as the calculated QCT reaction rate coefficients.   

\begin{acknowledgments}
This work has received funding from the {European Union's Horizon 2020 research and innovation program under the Marie Sklodowska-Curie} grant agreement no 894321. CMRR thanks also the Academic Leiden Interdisciplinary Cluster Environment (ALICE) provided by Leiden University for the computational resources. 
AJCV thanks the support of China’s Shandong Province ‘‘Double-Hundred Talent Plan’’ (2018), Coordenação de Aperfeiçoamento de Pessoal de N{\'i}vel Superior-Brasil (CAPES)-Finance Code 001, Conselho Nacional de Desenvolvimento Cient{\'i}fico e Tecnol{\'o}gico (CNPq), and Foundation for Science
and Technology, Portugal, in the framework of the project 55UIDB/00313/2020.
\end{acknowledgments}

\section*{Author Declarations}
\subsection*{Conflict of Interest}
The authors have no conflicts to disclose.

\section*{Data Availability Statement}
The full set of \emph{ab initio} grid points supporting the findings of this study is available from the corresponding author upon reasonable request. A Fortran subroutine of the final CHIPR PES that readily evaluates the potential and
gradient at any arbitrary geometry is made available as supplementary material. 


%

\end{document}